\documentclass{aa}

\usepackage{graphicx}
\usepackage{amsmath}
\usepackage{hyperref}
\usepackage{float}
\usepackage{placeins}
\usepackage[normalem]{ulem}

\usepackage{xcolor}
\usepackage{txfonts}

\usepackage{multirow}
\usepackage{svg}

\begin{document}

   \title{SN 2022xlp: The second-known well-observed, intermediate-luminosity
Iax supernova}

\titlerunning{SN 2022xlp: The second-known well-observed, intermediate-luminosity
Iax supernova}

   \author{D. Bánhidi\inst{1,2} \and
          B. Barna\inst{1} \and
          T. Szalai\inst{1} \and
          J. Vinkó\inst{1,3,5,14} \and
          I. B. Bíró\inst{2,4} \and
          K. A. Bostroem\inst{6} \and
          I.~Csányi\inst{2} \and
          K.~W.~Davis\inst{7} \and
          R.~J.~Foley\inst{7} \and
          L.~Galbany\inst{8,9} \and
          S. W. Jha\inst{10} \and
          D. A. Howell\inst{11,12} \and
          L. A. Kwok\inst{10} \and
          A. Pál\inst{3} \and
          C.~Pellegrino\inst{11,12} \and
          C.~Rojas-Bravo\inst{7} \and
          P.~Székely\inst{1} \and
          K.~Taggart\inst{7} \and
          G.~Terreran\inst{13} \and
          S.~Tinyanont\inst{15}
          }
          
   \institute{Department of Experimental Physics, Institute of Physics, University
of Szeged, D{\'o}m t{\'e}r 9, 6720 Szeged, Hungary \\ 
              \email{dbanhidi@titan.physx.u-szeged.hu}
        \and 
        Baja Astronomical Observatory of the University of Szeged, Szegedi
{\'u}t, Kt. 766, 6500 Baja, Hungary 
        \and
        HUN-REN CSFK Konkoly Observatory, Konkoly Thege M. út 15-17, Budapest,
1121, Hungary 
        \and
        HUN-REN-SZTE Stellar Astrophysics Research Group, Szegedi út, Kt.
766, 6500 Baja, Hungary 
        \and
        Department of Astronomy, University of Texas at Austin, 2515 Speedway,
Stop C1400, Austin, TX, 78712-1205, USA 
        \and
        Steward Observatory, University of Arizona, 933 North Cherry Avenue,
Tucson, AZ 85721-0065, USA 
        \and 
        Department of Astronomy and Astrophysics, University of California,
Santa Cruz, CA 95064, USA 
        \and
        Institute of Space Sciences (ICE, CSIC), Campus UAB, Carrer de Can
Magrans, s/n, E-08193 Barcelona, Spain 
        \and
        Institut d’Estudis Espacials de Catalunya (IEEC), E-08034 Barcelona,
Spain 
        \and
        Department of Physics and Astronomy, Rutgers, the State University
of New Jersey, 136 Frelinghuysen Road, Piscataway, NJ 08854-8019, USA 
        \and
        Las Cumbres Observatory, 6740 Cortona Drive, Suite 102, Goleta, CA
93117-5575, USA 
        \and
        Department of Physics, University of California, Santa Barbara, CA
93106-9530, USA 
        \and
        Center for Interdisciplinary Exploration and Research in Astrophysics
(CIERA), and Department of Physics and Astronomy, Northwestern University,
Evanston, IL 60208, USA 
        \and
        ELTE Eötvös Loránd University, Institute of Physics and Astronomy,
Pázmány Péter sétány 1A, Budapest, 1117, Hungary 
        \and
        National Astronomical Research Institute of Thailand (NARIT), Chiang
Mai, 50180, Thailand 
        }

   \date{Received 28/01/2025; accepted 18/07/2025}

  \abstract 
   {We present a detailed multicolor photometric and spectroscopic analysis
of type Iax supernova SN 2022xlp. With a V-band absolute magnitude light
curve peaking at $M_{max}(V) = -16.04 \pm 0.25\,\mathrm{mag}$, this object
is regarded as the second determined well-observed Iax supernova in the intermediate
luminosity range after SN 2019muj.
   }
   {Our research aims to explore the question of whether the physical properties
vary continuously across the entire luminosity range. We also investigate
 the chemical abundance profiles and the characteristic physical quantities
of the ejecta, followed by  tests of the predictions of hydro simulations.
   }
   {
   The pseudo-bolometric light curve was calculated using optical (BgVriz)
and UV (Swift UVOT UVW2,UVM2, UVW1,U,B) light curves and fits with a radiation
diffusion Arnett model to constrain the average optical opacity,  ejected
mass, and  initial nickel mass produced in the explosion. We analyzed the
color evolution of SN 2022xlp and compared it with that of other Iax supernovae
with different peak luminosities. We used the spectral tomography method
to determine the radial profiles of physical properties and abundances of
the ejecta, comparing them with a set of hydrodynamic pure deflagration models.
 
   }
   {SN 2022xlp shows a relatively rapid color evolution due to the decreasing
photospheric temperature in the early phase. The estimated bolometric flux
peaks at $8.87\times 10^{41}\,\mathrm{erg\,s^{-1}}$ and indicates the production
of radioactive nickel as $M(^{56}$Ni) = $0.0215 \pm 0.009\,M_{\odot}$. According
to the best-fit model, the explosion energy is $(2.066 \pm 0.236) \times
10^{49}\,\mathrm{erg}$ and the ejecta mass is $0.142 \pm 0.015\,\mathrm{M_{\odot}}$.
The performed spectral tomography analysis shows that the determined physical
quantities agree well with the predictions of the deflagration simulations,
with modifications regarding the increased Na abundance and the more massive
outer layers. SN 2022xlp bridges the previously existing luminosity gap,
together with SN 2019muj, and supports the assumption of continuous variation
in the physical properties across the SN Iax subclass.
   }
   {}


   \maketitle

\section{Introduction}
\label{section:introduction}

It is widely accepted that thermonuclear supernovae (SNe) are fusion explosions
of white dwarfs (WDs) in special binary systems. While modern observations
and theoretical studies, such as radiative energy transfer modeling in SN
ejecta and hydrodynamic simulations of explosions, have allowed us to unveil
the nature and properties of thermonuclear SNe, some key questions remain
unresolved. Over the past few decades, transient survey programs have discovered
several subtypes of thermonuclear events besides the Branch-normal SNe Ia
\citet{Branch-normal-Ia}. Currently, one of the most important questions
pertains to  exactly how the fusion explosion of a WD occurs and what the
progenitor systems are for the different types of thermonuclear SNe. In this
study, we present a detailed multicolor photometric and spectroscopic analysis
of a special thermonuclear supernova, SN 2022xlp, which belongs to the Iax
subclass.

The spectra of SNe Iax (often referred to as 2002cx-like objects after the
first known SN of this type of these peculiar transients \citep{SN2002cx} are similar to those of normal SNe Ia in terms of line-forming ions. SNe Iax always have
lower peak luminosity, lower photospheric velocities, shorter rise times,
and kinetic energies. Furthermore, these observables present a strong diversity
in the subclass. SN~2019gsc, one of the least luminous SN Iax, had a peak
brightness of $M_{max}(V)=-13.99\,\mathrm{mag}$ in the V-band \citep{SN2019gsc};
whereas SN~2011ay reached a peak at $M_{max}(V)=-18.39\,\mathrm{mag}$ \citep{SN2011ay-1}.
The rise time of SNe Iax typically falls between $10$ and $15$ days (see,
e.g., \cite{SN2008ha}, \cite{SN2019gsc}, and \cite{SN2020udy}), while the
decay phase of their light curves (LCs) is faster than that of the normal
SNe Ia.
In the case of SNe Iax, the value of the parameter $\Delta m_{15}$ (the decrease
in brightness within the first 15 days following the peak) typically falls
between $\sim 0.6 - 1.3\,\mathrm{mag}$. Finally, both the photospheric velocity
and the mass of the produced $^{56}\mathrm{Ni}$ isotopes display a broad
diversity; for instance, $v_\textrm{phot} = 2500\,\mathrm{km/s}$ and $M_{^{56}\mathrm{Ni}}
= 0.003\,\mathrm{M_{\odot}}$ were found in the case of SN 2008ha \citep{SN2008ha},
while the corresponding values for SN 2011ay, for instance, are $v_\textrm{phot}
= 9300\,\mathrm{km/s}$ and $M_{^{56}\mathrm{Ni}} = 0.225\,\mathrm{M_{\odot}}$
\citep{SN2011ay-1}. 

The spectral features of SNe Iax and normal SNe Ia also display significant
differences in their evolution. In SNe Iax, i) the lines of the intermediate
mass elements (IMEs) are weaker than those of the iron group elements (IGEs);
ii) the absorption components of the P Cygni line profiles cover shorter
range of wavelength, indicating that the line-forming processes of the particular
ions take part in a narrow atmospheric shell (caused by either the steeper
density or temperature gradient). At later epochs, SNe Iax never enter the
pure nebular phase; while some dominant forbidden emission lines are present,
the late-time spectra show continuum flux and permitted P Cygni profiles
(e.g. Fe\,II, Na\,I\,D, Ca\,II) even at more than $\sim 100\,\mathrm{days}$
after explosion \citep{Iax-spec-late}. These properties possibly indicate
that a bound remnant is left over after the unsuccessful disruption of a
WD, which is assumed to produce outflow by the super-Eddington wind process
\citep{Iax-Remnant}.

The most successful hydrodynamic simulations, for instance, reported in the
studies of \citet{Iax-def-Fink}, \citet{Iax-def-Lach}, \citet{Iax-def-Kromer},
and \citet{Iax-def-Min} have also shown that the pure deflagration explosion
of a WD is sufficient to explain the diversity of the observed properties
of SNe Iax. The strength of the initial deflagration determines the kinetic
energy and, thus, the expansion velocity of the ejecta. In contrast to the
detonation processes, the kinetic energy produced during pure deflagration
is too low to disrupt the whole WD. Thus, less material is ejected, producing
a lower-density profile compared to normal SNe Ia. The models predict that
the abundance profile of an SN Iax is almost constant and that IGEs have
higher abundances than in normal SNe Ia. According to these simulations,
the most abundant elements and isotopes are $^{56}\mathrm{Ni}$, $\mathrm{O,}$
and $\mathrm{C}$. 

In the case of typical SNe Ia, a relation exists between the peak absolute
brightness and the $\Delta m_{15}$ parameter \citep[usually called the Phillips
relation,][]{Phillips-re-Pskovskii,Phillips-rel-Phillips}, which allows for
the use of SNe Ia as precise distance indicators. Although SNe Iax basically
do not follow this relation \citep{Iax-SNe,HNB-SNe}, the question arises
of whether there are similar correlations regarding some of their physical
parameters. Previously, there have been attempts to find such pairs of correlated
parameters, such as the absolute magnitude of the peak in the V band and
the photospheric velocity, or the $\Delta m_{15}(V)$ parameter and the photospheric
velocity (\cite{Iax-relations-1}, \cite{SN14ck-Iax-relations-2}). These studies,
however, could not unambiguously determine the existence of such correlations
due to the limited sample size and the existence of outliers, such as SNe
2009ku \citep{SN2009ku}, or 2014ck \citep{SN14ck-Iax-relations-2}.

SN 2022xlp is considered special even within the Iax subclass because it
is only the second well-observed in the intermediate-luminosity range \citep[after
 SN~2019muj,][]{SN2019muj}. We present the context of the discovery of SN
2022xlp, along with the main properties of the host galaxy NGC 3938 in Section
\ref{section:SN2022xlp-NGC3938}. The observations, data reduction procedures,
and steps for detailed photometric and spectroscopic analysis are presented
in Sections \ref{section:obs-and-dataproc}, \ref{section:photmetry-photometric-analysis},
and \ref{section:spectroscopy-analysis-and-spectral-tomography}, respectively.
We summarize the results and conclusions of our study in Section \ref{section:summary}.

\section{SN 2022xlp and its host galaxy NGC 3938}
\label{section:SN2022xlp-NGC3938}
\begin{figure}
    \centering
    \includegraphics[width=250px]{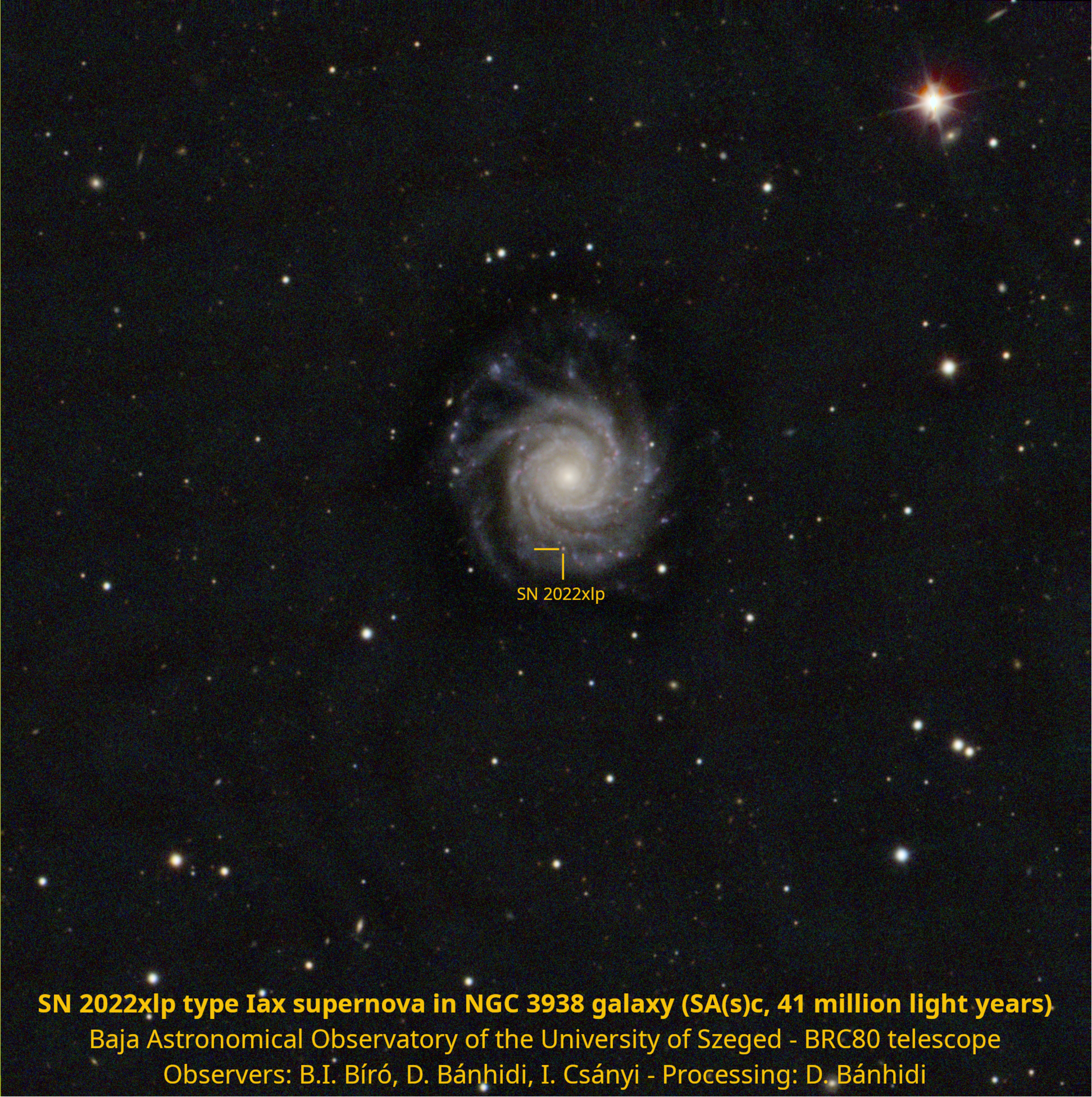}
    \caption{Image of NGC 3938 hosted the SN 2022xlp. This picture was taken
with the BRC80 telescope in the Baja Astronomical Observatory of the University
of Szeged.}
    \label{fig:NGC3938-SN2022xlp-BRC80-img}
\end{figure}

SN 2022xlp was discovered by Koichi Itagaki on $T_{0} = 59865.808\,\mathrm{MJD}$
with $\sim 17\,\mathrm{mag}$ in apparent magnitude \citep{22xlp-discovery}.
The supernova exploded in the NGC 3938 galaxy at RA = 11:52:49.560, Dec =
+44:06:03.60 (J2000). The redshift and the distance modulus of the host galaxy
are adopted from the NASA Extragalactic Database (NED\footnote{NED can be
accessed at \url{https://ned.ipac.caltech.edu/}}) and are listed in Table
$\ref{tab:NGC3938-SN2022xlp-data}$. The distance of NGC 3938 was measured
with the expanding photosphere method (EPM) based on the data analysis of
Type IIP SN 2005ay (which appeared earlier in this galaxy); the result in
the I-band is $d_{I} = 21.9\,\mathrm{Mpc,}$ while in the V-band $d_{V} =
22.4\,\mathrm{Mpc}$ \citep{dist-NGC3938}. For further analysis, we used the
average of these values $d = 22.15\,\mathrm{Mpc}$ as the distance from the
host galaxy. The galactic extinction data in the direction of NGC 3938 were
adopted from NED and listed in Table \ref{tab:NGC3938-SN2022xlp-data}. The
galactic color excess is $\mathrm{E(B}-\mathrm{V)} = 0.019\,\mathrm{mag}$.
SN 2022xlp was classified as SN Iax by Kenta Taguchi, Keiichi Maeda, and
Miho Kawabata (Kyoto University) based on the first spectrum obtained at
59867.8 MJD \citep{22xlp-classification}. The image of NGC 3938 with SN 2022xlp
can be seen in Fig. \ref{fig:NGC3938-SN2022xlp-BRC80-img}, while the detailed
data of these objects are listed in Table \ref{tab:NGC3938-SN2022xlp-data}.

\begin{table}
\caption{Data pertaining to NGC 3938 and SN 2022xlp. The NGC 3938 data were
adopted from NED, while the photometric data for SN 2022xlp were calculated
during the photometric analysis. See Section \ref{section:photmetry-photometric-analysis}
for details.}
\label{tab:NGC3938-SN2022xlp-data}
\centering
\begin{tabular}{ll}
\hline
Host galaxy                           & NGC 3938                        
          \\
Galaxy type                           & SA(s)c                          
          \\
Redshift                              & 0.002695                        
          \\
Heliocentric radial velocity          & $808 \pm 2\,\mathrm{km/s}$      
          \\
Distance modulus (without extinction) & $31.75 \pm 0.24\,\mathrm{mag}$  
          \\
Color excess ($E(B-V)$)               & $0.019\,\mathrm{mag}$           
          \\ \hline
Type of SN                            & Iax                             
          \\
Ra (J2000.0)                          & 11h 52m 49.45s                  
          \\
Dec (J2000.0)                         & +44$^{o}$ 07' 14.6"             
          \\ \hline
Estimated time of explosion           & $59861.8 \pm 0.5\,\mathrm{MJD}$ 
          \\[5pt]
Time of first detection               & $59865.8076\,\mathrm{MJD}$      
          \\[5pt]
Time of B-band maximum                & $59872.49 ^{+0.24}_{-0.18}\,\mathrm{MJD}$
 \\[5pt]
Time of V-band maximum                & $59874.06 ^{+0.20}_{-0.06}\,\mathrm{MJD}$
 \\[5pt] \hline
\end{tabular}
\end{table}

\newpage
\section{Observations and data processing}
\label{section:obs-and-dataproc}

\subsection{Photometry}
\label{subsection:photometry-obs-and-proc}
The long-term multicolor photometric follow-up of the supernova was started
at the Baja Astronomical Observatory of the University of Szeged with the
BRC80 telescope right after the classification, during the rising phase.
The instrument is a 0.8m Ritchey-Chrétien-Nasmyth (RCN) telescope, with
a focal length of 5700 mm and an f/7 light-gathering power. The telescope
is equipped with Johnson-Cousins-Bessel \textit{BV} and Sloan \textit{ugriz}
filters mounted in front of an FLI PL230-42 CCD camera, which has a 2048
$\times$ 2048 sensitive back-illuminated E2V CCD chip. The image scale is
0.55~arcsec/px, totaling to a field of view of 18.86$\times$18.86$\arcmin$.
We carried out standard multicolor photometric observations in 17 nights
between $-$5  and 150 days relative to the V-band maximum. The exposure times
are listed in the Table \ref{tab:BRC80-exptimes}. The instrumental magnitudes
of the SN were calculated using the image subtraction method and transformed
into the standard photometric system (for further technical details, see
Appendix \ref{appendix:BRC80-dataproc}).

\begin{figure}
    \centering    
    \includegraphics[width=1\linewidth]{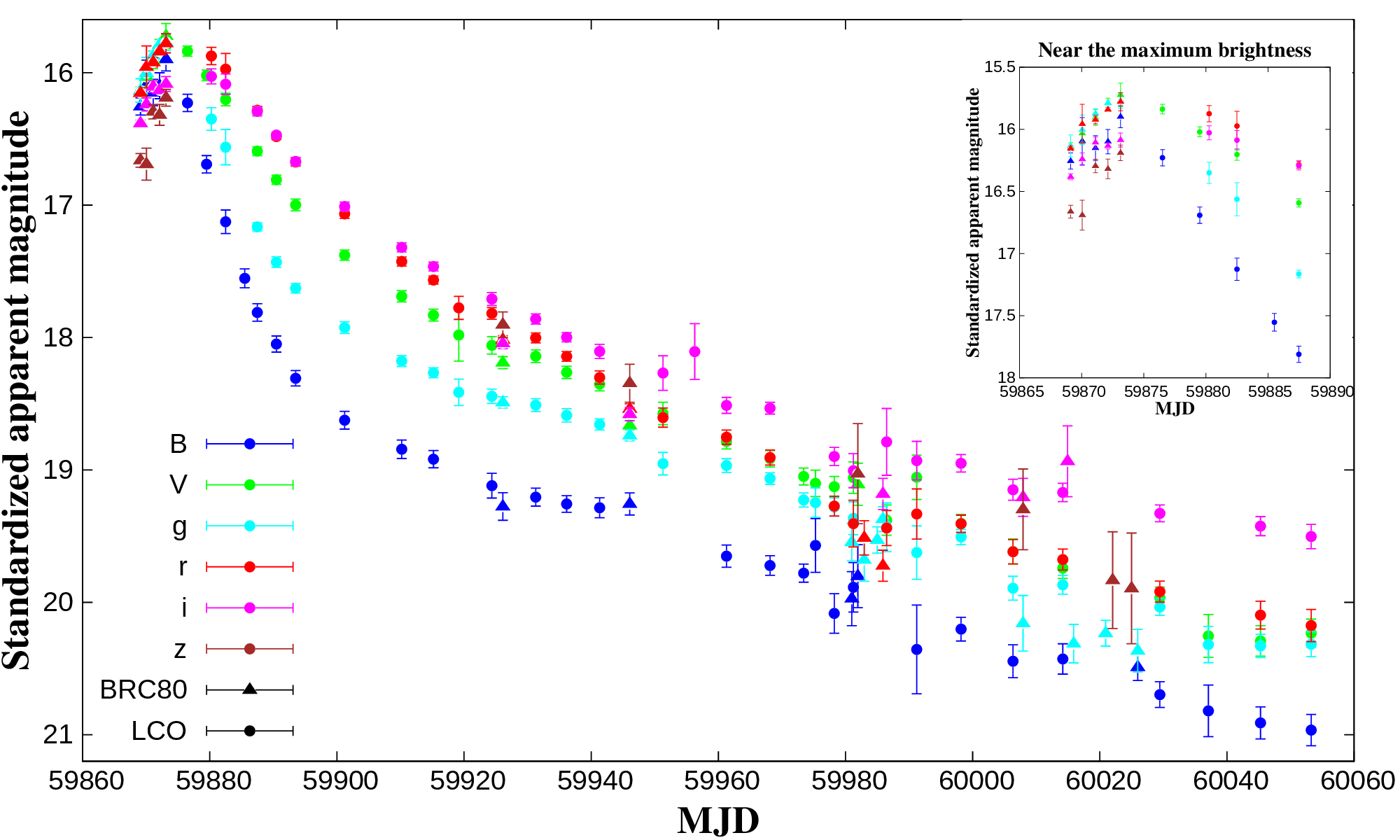}
    \caption{Ground-based photomety of SN 2022xlp. The zoom-inset highlights
the near-maximum observations. Data measured by the BRC80 telescope are marked
with triangles, while filled circles mark the LCO data points.}
    \label{fig:SN2022xlp-photometry}
\end{figure}

After maximum light, ground-based photometric follow-up was also obtained
with the 1m telescopes of the Las Cumbres Observatory (LCO\footnote{The official
website of the LCO can be accessed at \url{https://lco.global/}}) network
located at the Teide Observatory (Tenerife, Canary Islands, Spain) and McDonald
Observatory (Fort Davis, Texas, USA) via the Global Supernova Project (GSP)
in the SDSS \textit{gri} and Johnson-Cousins-Bessel \textit{BV} bands.
The data obtained by the BRC80 telescope were reduced by our IRAF-based reduction
and photometric software, while the median
combination and the aperture photometry of LCO data were performed by applying
self-written pipelines using the {\tt FITSH} software \citet{FITSH}. The
long-term multicolor standard photometric magnitudes of the SN 2022xlp are
listed in Table \ref{tab:SN2022xlp-photometry}, while the standard multicolor
photometric LCs are plotted in Fig. \ref{fig:SN2022xlp-photometry}. For the
epoch with no available measurements in the Johnson-Cousins-Bessel B and
V bands at LCO, these magnitudes were calculated by transformation of SDSS
band magnitudes according to \citet{SDSS-to-Bessel}, while data points with
errors greater than $\sim0.1\,\mathrm{mag}$ were rejected.

In addition to ground-based photometry, the Neil Gehrels Swift Observatory
Ultraviolet/Optical Telescope (UVOT, hereafter Swift; \citet{Swift-1}; \citet{Swift-2};
\citet{Swift-3}) measured the brightness in the UV-band filters (W2, M2,
W1, U, B) at three epochs between -4 and 3 days relative to the V-band maximum.
The Swift data were downloaded from the Swift archive. These data were reduced
using standard HEAsoft tasks. Individual frames were summarized with the
{\tt uvotimsum} task. The magnitudes were determined via aperture photometry
using the task {\tt uvotsource} and adopting the most recent zero-points
from the {\tt caldb} database. The UV photometric dataset is presented in
Table \ref{tab:SN2022xlp-UV-photometry}.

\subsection{Spectroscopy}
\label{subsection:spectrocopy-obs-and-proc}
The first spectrum was taken with the KOOLS-IFU spectrograph attached to
the 3.8-m Seimei telescope at the Okayama Observatory. After the classification,
a spectroscopic follow-up was obtained with the 3m Shane telescope located
in the Lick Observatory of the University of California. The spectra were
observed with the Shane telescope's Kast Dual Channel Spectrograph \citep{3mShane-KAST}.
The telescope's three-meter primary mirror and the spectrograph's efficient
light throughput enabled the recording of spectra with a high signal-to-noise
ratio (S/N), required by spectral tomography. To reduce the Kast data, we
used the {\tt UCSC Spectral Pipeline}\footnote{\url{https://github.com/msiebert1/UCSC\_spectral\_pipeline}}
\citep{Siebert20} data reduction pipeline based on procedures outlined by
\citet{Foley03}, \citet{Silverman2012}, and references therein.  The two-dimensional
(2D) spectra were bias-corrected, flat-field corrected, adjusted for varying
gains across different chips and amplifiers, and trimmed.  The one-dimensional
(1D) spectra were extracted using the optimal algorithm \citep{Horne86}.
 The spectra were wavelength-calibrated using internal comparison-lamp spectra
with linear shifts applied by cross-correlating the observed night-sky lines
in each spectrum to a master night-sky spectrum.  Flux calibration and telluric
correction were performed using standard stars at an air mass similar to
that of the science exposures. We combined the sides by scaling one spectrum
to match the flux of the other in the overlap region and using their error
spectra to correctly weight the spectra when combining.  More details of
this process are discussed in prior works \citep{Foley03, Silverman2012,
Siebert20}. These observations were primarily coordinated using {\tt YSE-PZ}
\citep{Coulter22, Coulter23}, an open source, general-purpose target and
observation management (TOM) platform.  

From +23 days since the V-band maximum, the 2m Faulkes Telescope North of
the Las Cumbres Observatory (LCO\footnote{The official website of the LCO
can be accessed at \url{https://lco.global/}}) also joined the spectral follow-up
through the Global Supernova Project. Spectra were taken with the FLOYDS
spectrograph\footnote{Detailed description of the instrument can be found
at \url{https://lco.global/observatory/instruments/floyds/}}. The 1D spectra
were extracted using the FLOYDS pipeline\footnote{Detailed description of
FLOYDS pipeline is available at \url{https://github.com/svalenti/FLOYDS_pipeline}}.

\begin{figure*}
    \centering
    \includegraphics[width=0.94\linewidth]{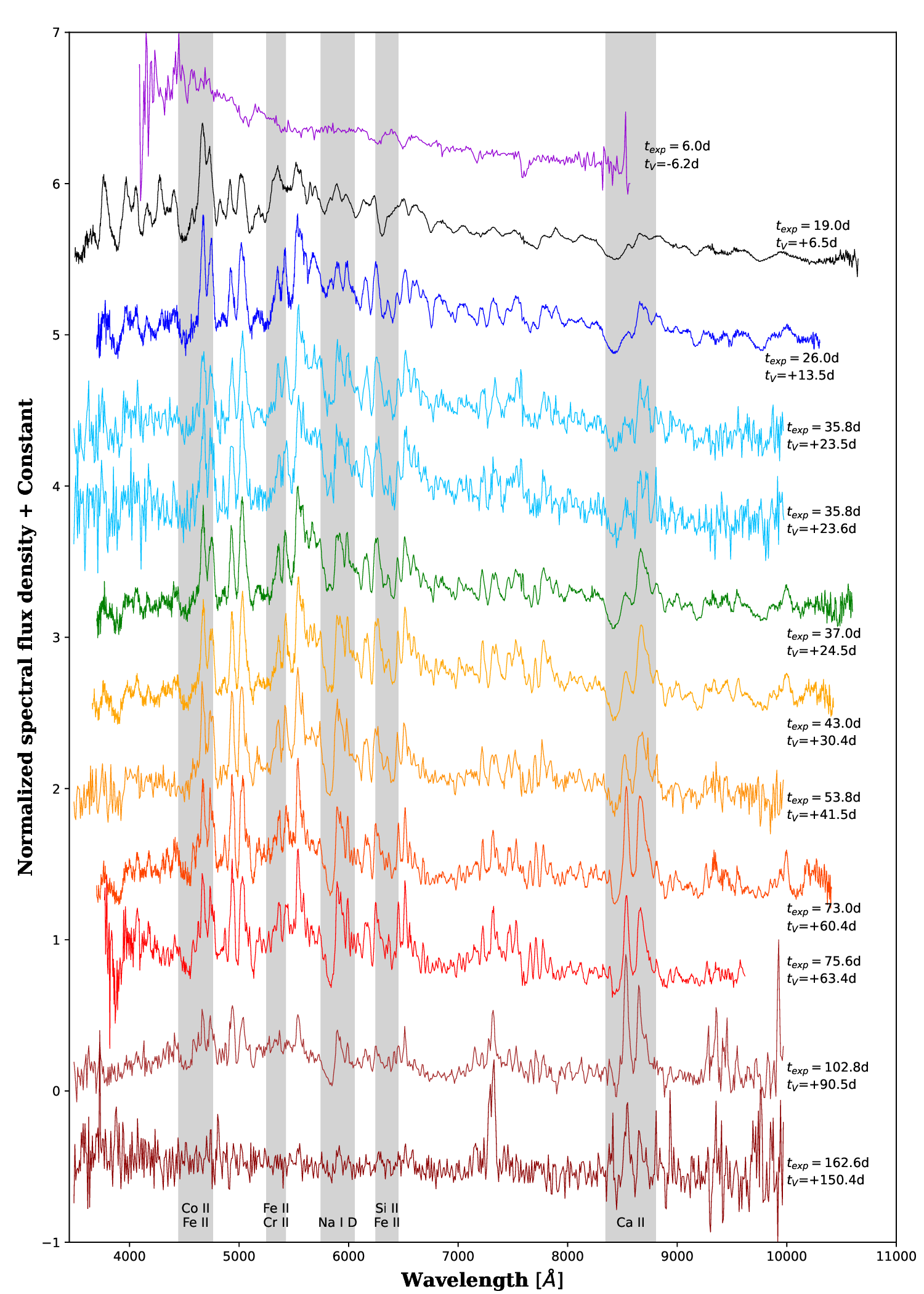}
    \caption{Optical spectra of SN 2022xlp. The characteristic spectral features
are highlighted by gray stripes. The phase relative to the V-band maximum
($t_\textrm{V}$) and the time since the explosion ($t_\textrm{exp}$) are
also marked.}
    \label{fig:SN2022xlp-VIS-norm-spectra-series}
\end{figure*}

In the case of the classification spectrum and the spectra taken by the 2m
Faulkes telescope, we applied a moving-averaging process to improve the S/N.
We performed a wavelength-dependent spectral flux calibration on all spectra
according to match the multicolor photometry. 
The dataset was corrected for extinction, reddening, and redshift with values
adopted from the NED database (see Table \ref{tab:NGC3938-SN2022xlp-data}).
In the absence of a high-resolution spectrum, it was not possible to precisely
measure the extinction of the host galaxy using the Na\,I\,D lines. Since
the explosion occurred in a region of the NGC 3938 disk with less interstellar
matter, viewed face-on, we neglected the local extinction and only applied
a correction for the Galactic component.
The series of normalized spectra can be seen in Fig. \ref{fig:SN2022xlp-VIS-norm-spectra-series},
where  the main spectral features are also highlighted. The log of spectral
observations is summarized in Table \ref{tab:SN2022xlp-spectra-data}.

\FloatBarrier
\clearpage

\section{Photometry and light curve analysis}
\label{section:photmetry-photometric-analysis}

We fit individual LCs with simple polynomial functions to determine the characteristic
parameters, namely the MJD of the peak brightness, $\Delta m_{15}$, and the
apparent and absolute magnitudes of the peak. Hereafter, we use the moment
of the V-band maximum as a reference time, since most of the data points
are available in this filter. The determined LC parameters are listed in
Table \ref{tab:SN2022xlp-photometric-params}. According to the peak absolute
brightnesses of the $M_{\mathrm{peak}}(V) = -16.04 \pm 0.25\,\mathrm{mag}$
and $M_{\mathrm{peak}}(g) = -16.05 \pm 0.25\,\mathrm{mag}$, SN 2022xlp is
the second intermediate luminous supernova studied in detail along with SN
2019muj ($M_{\mathrm{peak}}(V) = -16.42 \pm 0.06\,\mathrm{mag}$ and $M_{\mathrm{peak}}(g)
= -16.48 \pm 0.07\,\mathrm{mag}$) \citep{SN2019muj}. The obtained decline
rates are $\Delta m_{15}\mathrm{(V)} = 0.94 ^{+0.04}_{-0.02}\,\mathrm{mag}$
and $\Delta m_{15}\mathrm{(g)} = 1.48 ^{+0.02}_{-0.02}\,\mathrm{mag}$ indicating
a slower dimming than that of SN 2019muj \citep[$\Delta m_{15}\mathrm{(V)}
= 1.2\,\mathrm{mag}$ and $\Delta m_{15}\mathrm{(g)} = 2.0\,\mathrm{mag}$][]{SN2019muj}.

\begin{table*}
\label{tab:SN2022xlp-photometric-params}
\caption{Main photometric parameters of the multicolor LCs of SN 2022xlp.
The MJD of peak brightness, $\Delta m_{15}$ parameter, peak apparent magnitude
and peak absolute magnitude are listed for the Johnson-Cousins-Bessel B,V
and SDSS g,r,i and z bands. These parameters were obtained via the polynomial
fitting of the LCs by {\tt scipy.optimize.curve\_fit}.}
\centering
\begin{tabular}{ccccc}
\hline
Filter & MJD of peak brightness & $\Delta m_{15}\,\mathrm{[mag]}$ & $m_{max}\,\mathrm{[mag]}$
& $M_{max}\,\mathrm{[mag]}$ \\[1pt] \hline
B &  $59872.49 ^{+0.24}_{-0.18}$  &  $1.79 ^{+0.01}_{-0.02}$  &  $16.00 ^{+0.11}_{-0.09}$
 &  $-15.83 ^{+0.27}_{-0.26}$   \\[5pt] 
V & $59874.06 ^{+0.20}_{-0.06} $  &  $0.94 ^{+0.04}_{-0.02}$  &  $15.76 ^{+0.05}_{-0.08}$
 &  $-16.04 ^{+0.25}_{-0.25}$   \\[5pt] 
g & $59873.28 ^{+0.12}_{-0.24} $  &  $1.48 ^{+0.02}_{-0.02}$  &  $15.77 ^{+0.05}_{-0.04}$
 &  $-16.05 ^{+0.25}_{-0.24}$   \\[5pt] 
r & $ 59875.64 ^{+0.70}_{-1.24}$  &  $0.75 ^{+0.07}_{-0.10}$  &  $15.74 ^{+0.07}_{-0.07}$
 &  $-16.06 ^{+0.25}_{-0.25}$   \\[5pt] 
i & $59876.96 ^{+0.56}_{-1.28} $  & $0.57 ^{+0.06}_{-0.10}$   &  $15.99 ^{+0.07}_{-0.07}$
 &  $-15.80 ^{+0.25}_{-0.25}$   \\[1pt]
\hline \hline
\end{tabular}
\end{table*}

\begin{figure}[H]
    \centering
    \includegraphics[width=0.90\linewidth]{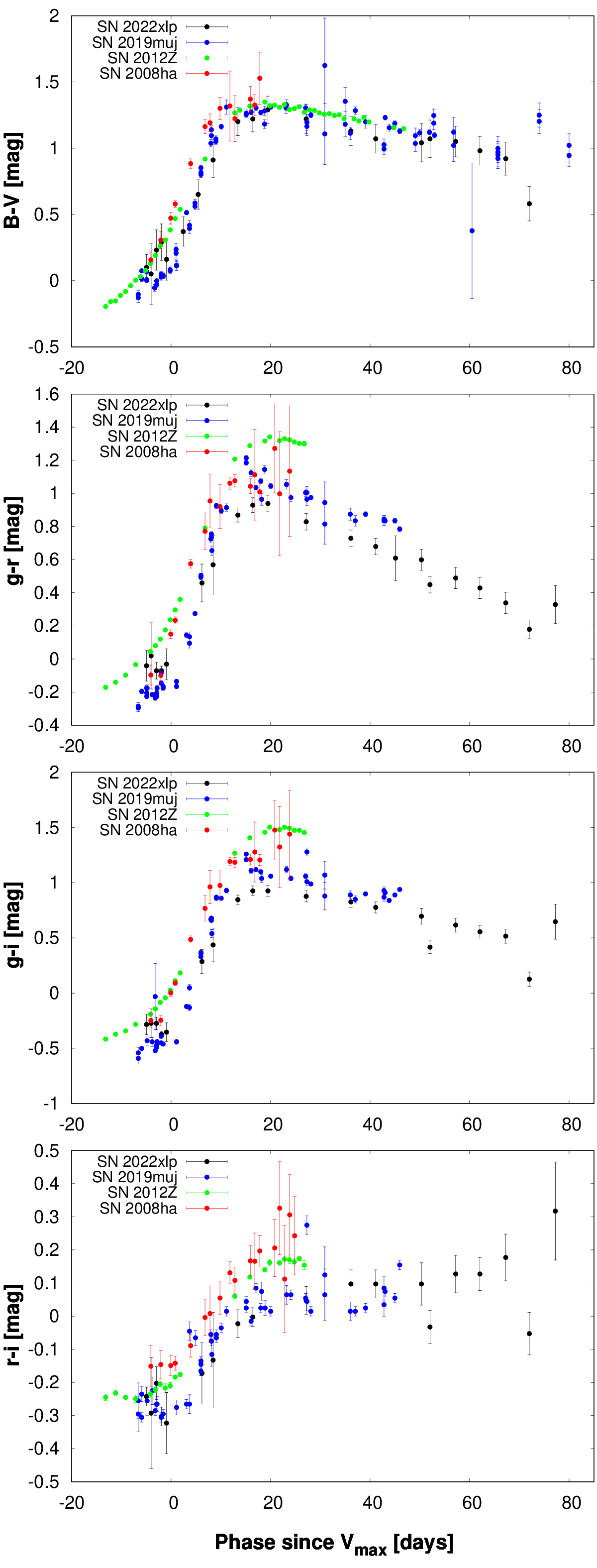}
    \caption{Evolution of different color indexes compared to another SNe
Iax (SN 2019muj - \citep{SN2019muj}, SN 2012Z - \citep{SN2012Z}, and SN 2008ha
- \citep{SN2008ha};\citep{SN2008ha-SN2010ae-2}). SN 2022xlp shows a very
similar color evolution to SN 2019muj. The color indexes are corrected for
the extinction as described in Sect. \ref{section:photmetry-photometric-analysis}.
We adopted the color excess from the papers listed above.}
    \label{fig:SN2022xlp-color-evolution}
\end{figure}

The color evolution of SN 2022xlp can be seen in Fig. \ref{fig:SN2022xlp-color-evolution},
along with a comparison to other Iax supernovae. We applied extinction corrections
to the color indexes with the formula $(m_{\lambda 1} - m_{\lambda 2})_{real}
= (m_{\lambda 1} - m_{\lambda 2})_{obs} - (R_{\lambda 1} - R_{\lambda 2})\times
E(B-V)$, where $(m_{\lambda 1} - m_{\lambda 2})_{real}$ is the extinction-corrected
color index, while $(m_{\lambda 1} - m_{\lambda 2})_{obs}$ is the observed
color index, $E(B-V)$ is the color excess, and $R_{\lambda}$ is a filter-dependent
coefficient. The time dependence of the various color indices shows a very
strong similarity in the case of SN 2022xlp and SN 2019muj. According to
the B-V, g-r, and g-i color changes, a relatively steep reddening occurs
in the very early phase between about $-5\,\mathrm{and}\,18\,\mathrm{days}$
and the V-band maximum. The B-V color index changes $\sim 1.3\,\mathrm{mag}$
in $\sim 23\,\mathrm{days}$, while the g-r and g-i color indexes vary $\sim
1.0\,\mathrm{mag}$ and $\sim 1.2\,\mathrm{mag,}$ respectively. This reddening
is much faster compared to normal SNe Ia \citep{Iax-reddening}. The r-i color
shows a monotonic increase, which becomes quasi-linear after 16 days following
the V maximum. In Fig. \ref{fig:SN2022xlp-color-evolution},  the color evolution
of SN 2022xlp is compared to a fainter and brighter SNe Iax, SN 2008ha and
SN 2012Z. All SNe Iax show an overall similar color evolution with minor
differences in the amplitude and steepness of the initial reddening. The
fainter SNe Iax shows a faster color change rate. The amplitude of the color
change is greater in the case of brighter Iax supernovae. After the reddening
maximum, the color indices start to decline slowly and quasi-linearly.

A pseudo-bolometric LC is created by the integration of spectral flux densities
calculated from the multicolor magnitudes. The optical flux is calculated
with the {\tt SuperBol} Python software \citep{SuperBol}. We calculate the
UV flux with the Swift UVOT magnitudes. At every third epoch, the magnitudes
were corrected to extinction, converted to spectral flux densities, and integrated
up to the effective wavelength of the B filter. According to these three
epochs, the time evolution of the UV flux follows a linear decay trend. We
fit the $F_{UV}-t_{V}$ data points with a linear function to calculate the
UV part of the total flux for other epochs when only optical data are available.

The IR contribution was estimated from the \textit{i} band flux. We adopted
$F_{IR} = \lambda F_{\lambda}/3$ according to the integrated formula of the
Rayleigh-Jeans assumption, where $\lambda$ is the effective wavelength of
the \textit{i} filter.

We performed an analysis of the pseudo-bolometric LC via the fitting of the
theoretical (Arnett) model. The expansion of the supernova ejecta was assumed
to be spherically symmetric and homologous \citep{SN2015H}. At the pre-maximum
epochs, the ejecta can be modeled as an optically thick expanding fireball
model. The energy released by the decay of $^{56}\mathrm{Ni}$ (with half-life
time $t_{1/2}=6.077\,\mathrm{days}$ and the energy production rate $\epsilon_{^{56}\mathrm{Ni}}=3.9\times10^{10}\,\mathrm{erg/s/g}$)
is in agreement with the decreasing internal energy caused by the adiabatic
expansion and radiation. Thus, the temperature can remain nearly constant
and the luminosity of the supernova has a quadratic dependence on time, expressed
as
\begin{equation}
    L = 4 \pi \sigma R^2 T^4 = 4 \pi \sigma v_\mathrm{ph}^2 t_\mathrm{exp}^2
T_\mathrm{ph}^4 = \mathrm{constant}\times t_\mathrm{exp}^2
,\end{equation}
where $R$ is the radius of the supernova photosphere, $T_{ph}$ and $v_{ph}$
are its temperature and expansion velocity, and $t_{exp}$ is the time since
the explosion. At maximum light, the emitted energy is equal to the energy
released by the decay of $^{56}\mathrm{Ni}$; thus, the peak luminosity depends
on the nickel mass produced. The post-maximum phase of the LC is also influenced
by the decay of $^{56}\mathrm{Ni}$ and its daughter isotope $^{56}\mathrm{Co}$,
and the decline rate is proportional to the effective diffusion time of the
photons: $\Delta m_{15} \sim t_{d} \sim \left(  \frac{\kappa^2 M^3}{E_{k}}
\right)^{1/4}$, where $t_{d}$ is the diffusion time, $\kappa$ is the opacity,
$M$ is the mass of the supernova ejecta, and $E_{k}$ is its kinetic energy.
The opacity is mainly affected by the Thomson-scattering of photons on free
electrons and the fluorescence of IGEs. 

\begin{figure}[H]
    \centering
    \includegraphics[width=1\linewidth]{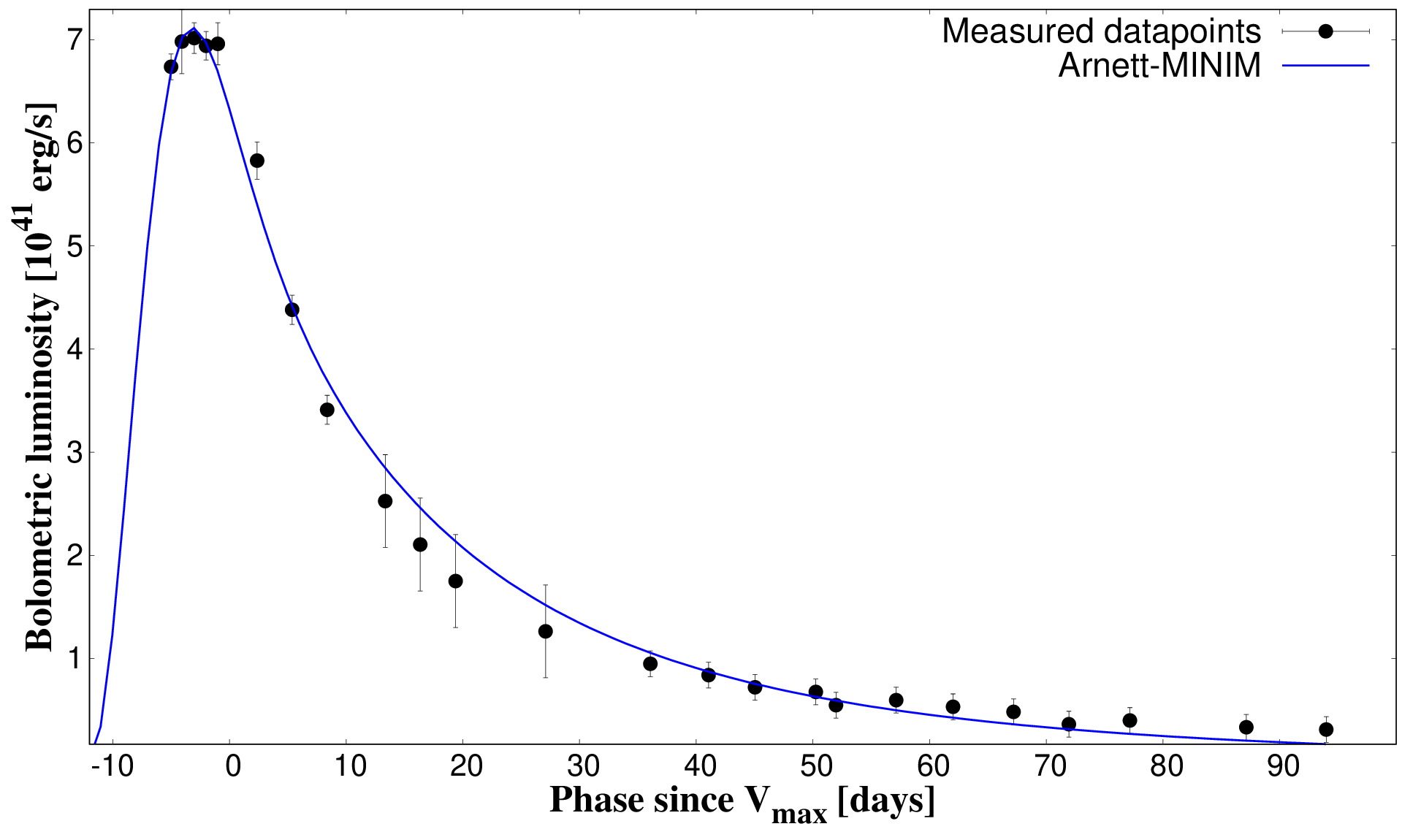}
    \caption{Pseudo-bolometric LC of SN 2022xlp and the best-fit synthetic
light model.}
    \label{fig:SN2022xlp-Arnett}
\end{figure}

The time-dependent bolometric flux can be expressed with a semi-analytic
model as presented in \citet{Arnett-MINIM-Vinko}, which is based on the radioactive
decay and photodiffusion model described in \cite{Arnett-1} and \cite{Arnett-Valenti}
as

\begin{multline}
L(t) = M_{^{56}Ni}\exp{(-x^2)}\left[1-\exp{\left(-\frac{t_{\gamma}}{t_{exp}^2}
\right)}\right] \times \\
\times \left[(\epsilon_{^{56}Ni}-\epsilon_{^{56}Co})\int_{0}^x 2z\exp{(z^2-2zy)}dz
+ \right.\\
\left. + \epsilon_{^{56}Co}\int_{0}^x 2z\exp{(z^2-2zy+2zs)}dz \right],
\label{eq:Arnett-model}
\end{multline}

where $t_{exp}$ is the time since the explosion, $M_{^{56}Ni}$ is the mass
of the $^{56}Ni$ produced by the explosion, and $\epsilon_{^{56}Ni}$ and
$\epsilon_{^{56}Co}$ are the energy production rates of the $^{56}Ni$ and
$^{56}Co$ isotopes, respectively. The factor $t_{\gamma} = (3\kappa_{\gamma}M_{ej})/(4\pi
v^2)$ accounts for the gamma leakage, where $k_{\gamma}$ is the gamma ray
opacity, $v$ is the expansion velocity, $M_{ej}$ is the mass of the ejecta.
The remaining parameters are defined by the following expressions: $x=t/t_d$,
$y=t_d/(2t_{Ni})$, $z=1/(2t_d),$ and $s=t_d(t_{Co}-t_{Ni})/(2t_{Co}t_{Ni})$,
where $t_d = \sqrt{(2 \kappa M_{SN})/(\beta c v)}$ is the effective diffusion
time, $\kappa$ is the total opacity and $\beta = 13.8$ is a fixed LC parameter,
which is related to the density distribution \citep{Arnett-1}. The synthetic
LC produced by the presented semi-analytic model is fit to the pseudo-bolometric
LC by the {\tt MINIM} code \citep{MINIM}. The best-fit model parameters are
listed in Table \ref{tab:SN2022xlp-Arnett-params}, while the pseudo-bolometric
LC and the best-fit synthetic model are shown in Fig. \ref{fig:SN2022xlp-Arnett}.
According to the determined Arnett model parameters, the total mass of the
ejecta and the explosion energy can be calculated with the following formulae:
\begin{equation}
    M_{ej} = \frac{4 \pi t_{\gamma}^2 v_{ph}^2}{3 \kappa_{\gamma}}
    \label{eq:ejecta-mass}
,\end{equation}
\begin{equation}
    E_{kin} = \frac{3}{10} M_{ej} v_{ph}^2
\label{eq:exp-energy}
,\end{equation}
where $v_{ph}$ is the photospheric velocity, $\kappa_{\gamma} = 0.025\,\mathrm{cm^2g^{-1}}$
according to \citep{kappa-gamma}. In the case of Eq. \ref{eq:exp-energy},
constant density ejecta is assumed. The photospheric velocity can be determined
by spectral analysis, which is described in Sect. \ref{section:spectroscopy-analysis-and-spectral-tomography}.
We can determine the mass and kinetic energy of the ejecta based on the effective
diffusion time with the formulae adopted from \citet{ejecta-trise}:

\begin{equation}
    M_{ej} = 0.66\,M_{\odot} \left( \frac{t_d}{16\,\mathrm{days}} \right)^2
\left( \frac{v_{ph}}{6000\,\mathrm{km\,s^{-1}}} \right)
\label{eq:ejecta-mass-from-td}
,\end{equation}

\begin{equation}
    E_{kin} = 2.4 \times 10^{50}\,\mathrm{erg} \left( \frac{t_d}{16\,\mathrm{days}}
\right)^2 \left( \frac{v_{ph}}{6000\,\mathrm{km\,s^{-1}}} \right)^3
\label{eq:exp-energy-from-td}
.\end{equation}

Regarding the determined model parameters, the mass of $^{56}Ni$ is comparable
and slightly lower than in the case of SN~2019muj, as predicted by similar
absolute peak brightnesses. The nickel mass determined from the Arnett model
fit is $M_{^{56}Ni} = 0.0191 \pm 0.0025\,\mathrm{M_{\odot}}$, while in the
case of SN~2019muj, it is $M_{^{56}Ni} = 0.031 \pm 0.005\,\mathrm{M_{\odot}}$
\citep{SN2019muj}. The effective diffusion time is similar for SN~2019muj
and SN~2022xlp, $7.1 \pm 0.5\,\mathrm{days}$ for SN 2019muj and $6.46 \pm
1.73\,\mathrm{days}$ for SN~2022xlp, while the gamma leakage timescale differs
significantly, with $5.5 \pm 1.5\,\mathrm{days}$ for the former and $19.26
\pm 5.23\,\mathrm{days}$ for the latter. The timescale of gamma-ray leakage
cannot be accurately estimated because it is most sensitive for the middle
part of the bolometric LC and the uncertainty is the largest at that part.
Thus, the ejecta parameters determined from the timescale of gamma-ray leakage
suffer a more significant error. 

The ejecta mass calculated from the gamma-ray leakage timescale is $M_{ej}
= 0.056 \pm 0.031\,M_{\odot}$, while from the effective diffusion timescale,
it is estimated to be $M_{ej} = 0.088\pm0.047\,M_{\odot}$. These results,
derived from independent model parameters, are in agreement within the margin
of error. In the case of SN~2019muj, the ejecta mass is determined to be
$M_{ej} = 0.17\,M_{\odot}$ \citet{SN2019muj}, which is slightly greater than
that of SN 2022xlp. 

Applying Eqs. \ref{eq:exp-energy} and \ref{eq:exp-energy-from-td} to determine
the kinetic energy of the explosion based on the timescale of gamma-ray leakage
and the effective diffusion time, we found $E_{kin} = (8.16 \pm 4.66) \times
10^{48}\,\mathrm{erg}$ and $E_{kin} = (2.16 \pm 3.31) \times 10^{49}\,\mathrm{erg}$.
These results are approximately of the same order of magnitude. The kinetic
energy of SN~2019muj was also found to be slightly higher, at $E_{kin} =
4.4\times10^{49}\,\mathrm{erg}$ \citet{SN2019muj}.

Based on Table \ref{tab:SN2022xlp-Arnett-params}, we can compare the results
with a brighter \citet[SN 2012Z,][]{SN2012Z} and a fainter (SN 2008ha \citet{SN2008ha};
\citet{SN2008ha-SN2010ae-2}) SN Iax. The brighter SN 2012Z shows an order
of magnitude higher Ni- and ejecta mass, while its kinetic energy differs
two orders of magnitude from the intermediate-luminosity SNe Iax. In the
case of the fainter one, SN 2008ha, the determined Ni mass and the kinetic
energy are one order of magnitude lower, while the ejecta mass is similar
to the intermediate-luminosity SNe Iax. We note that the ejecta mass can
often be determined only with relatively low precision between $0.02\,M_{\odot}$
and $0.2\,M_{\odot}$ and these uncertainties propagate to the kinetic energy
as well.

The determined ejecta parameters, namely, $E_{exp}$ as the explosion energy,
$M_{^{56}Ni}$ as the nickel mass, and $M_{ej}$ as the ejecta mass, can be
compared with the results of various hydrodynamic deflagration simulations;
for example, from \citet{Iax-def-Lach}. Table \ref{tab:SN2022xlp-Arnett-params}
also presents the parameters of the two deflagration models whose peak luminosities
are closest to that of SN 2022xlp. The determined ejecta parameters agree
well with the parameters of {\tt def\_2021\_r48\_d5.0\_z} model.

Based on the fitting of the bolometric LC model, the rise time is constrained
as $12.04\,\mathrm{days}$ in the V-band, marking the moment of the first
light at $\mathrm{T_0}=59862.02$ MJD. This is in good agreement with the
explosion date estimated from the spectral analysis (see Sect. \ref{section:spectroscopy-analysis-and-spectral-tomography}),
which is $-12.26\,\mathrm{days}$ ($\mathrm{T_{exp}}=59861.8$ MJD). However,
note that the two characteristic dates are not equivalent, since $\mathrm{T_{exp}}$
precedes $\mathrm{T_{0}}$, as the first photons are released after the excitation
with some delay.  The difference is expected to be a few hours to two days
for normal SNe Ia \citet{Ia-LC-rise}; therefore, the inferred dates of SN
2022xlp are consistent with each other.

\begin{table*}
\label{tab:SN2022xlp-Arnett-params}
\caption{Best-fit semi-analytic (Arnett) model parameters from the pseudo-bolometric
LC fitting and the properties calculated with these parameters in the case
of SN 2008ha \citep{SN2008ha}, SN 2022xlp, SN 2019muj \citep{SN2019muj} and
SN 2012Z \citep{SN2012Z}. In the case of SN 2022xlp, the * sign refers to
the ejecta parameters calculated from Equations \ref{eq:ejecta-mass} and
\ref{eq:exp-energy}, while ** refers to the results determined from Equations
\ref{eq:ejecta-mass-from-td} and \ref{eq:exp-energy-from-td}. The last two
rows present the results of two deflagration hydrodynamic simulations from
\citet{Iax-def-Lach}.}
\centering
\begin{tabular}{llllll}
\hline
SN & $M_{^{56}\mathrm{Ni}}\,\mathrm{[M_{\odot}]}$ & $t_{d}\,\mathrm{[days]}$
& $t_{\gamma}\,\mathrm{[days]}$ & $M_{ej}\,\mathrm{[M_{\odot}]}$ & $E_{exp}\,\mathrm{[10^{49}\,erg]}$
\\ \hline
SN 2008ha    & $0.003 \pm 0.001$   & -                 & -              
 & $0.15$            & $0.23$             \\
SN 2022xlp*  & $0.0191 \pm 0.0025$ & $6.463 \pm 1.703$ & $19.26 \pm 5.23$
& $0.056 \pm 0.031$ & $0.816 \pm 0.446$ \\
SN 2022xlp** & $0.0191 \pm 0.0025$ &  $6.463 \pm 1.703$ & $19.26 \pm 5.23$
& $0.088 \pm 0.047$  & $2.160 \pm 3.310$ \\
SN 2019muj   & $0.031 \pm 0.005$   & $7.1 \pm 0.5 $    & $5.5 \pm 1.5$  
 & $0.17$             & $4.4$             \\ 
SN 2012Z     & $0.25 \pm 0.02$     & -                 & -              
 & $2.0 \pm 0.6$        & $175 \pm 105$     \\ \hline
def\_2021\_r120\_d5.0\_z & 0.010 & - & - & 0.024 & 0.34 \\
def\_2021\_r48\_d5.0\_z & 0.018 & - & - & 0.054 & 0.72 \\
\hline \hline
\end{tabular}
\end{table*}

\FloatBarrier

\section{Spectral analysis and spectral tomography}
\label{section:spectroscopy-analysis-and-spectral-tomography}
\begin{figure*}
    \centering
    \includegraphics[width=1\linewidth]{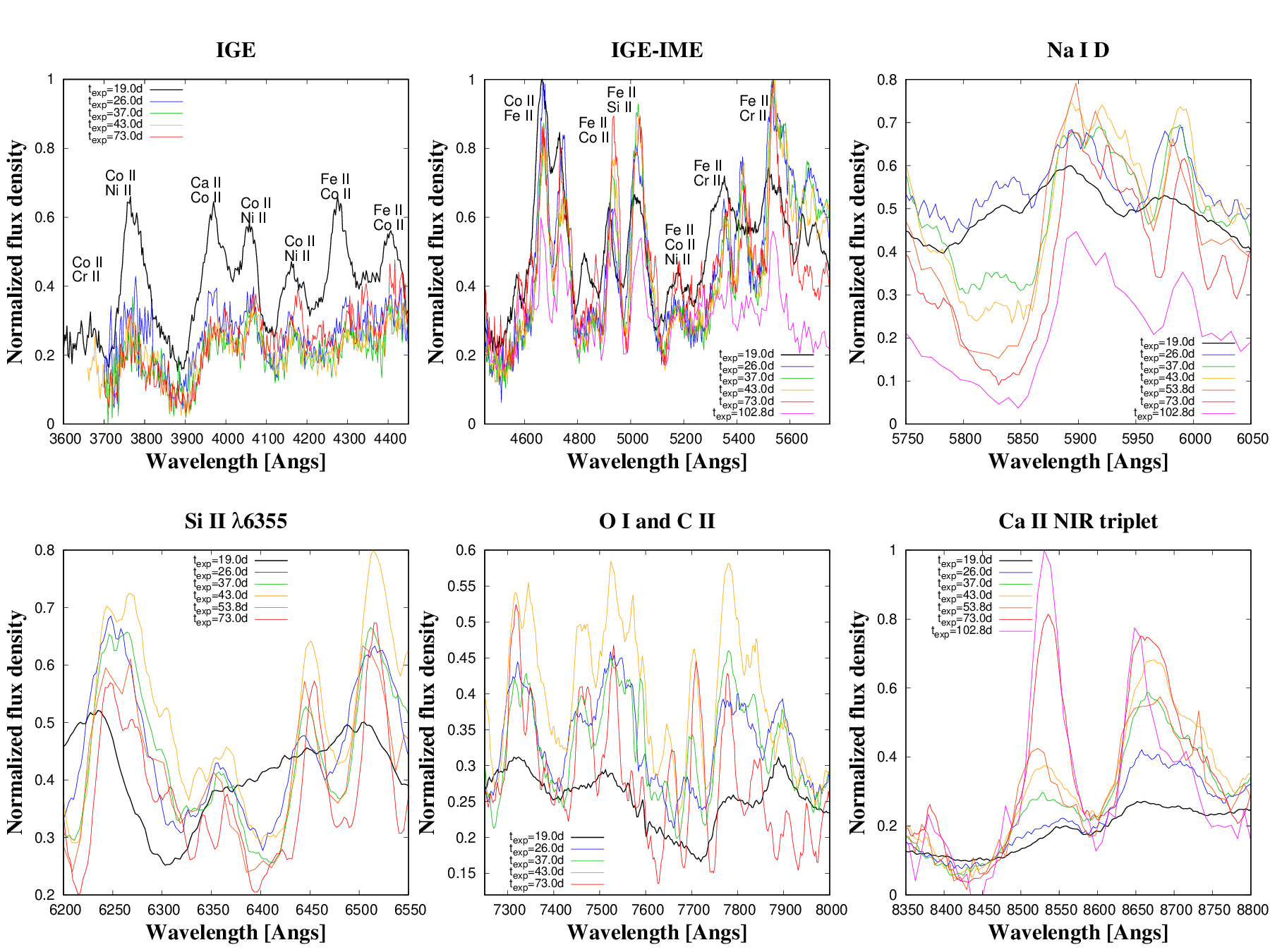}
    \caption{Zoom-in on the details of the normalized spectral time series
of SN 2022xlp shown in Fig. \ref{fig:SN2022xlp-VIS-norm-spectra-series} and
described in Table \ref{tab:SN2022xlp-spectra-data}, focusing on characteristic
spectral lines. The lines specified in the legend correspond to the different
epochs observed. Note: to improve clarity, not all epochs are plotted. }
    \label{fig:SN2022xlp-spectral-line-evolution}
\end{figure*}

We analyzed the evolution of the characteristic spectral lines of SN 2022xlp
by comparing its normalized spectra taken at various epochs, as shown in
Fig. \ref{fig:SN2022xlp-spectral-line-evolution}. At -6.2 days, the spectrum
shows only a few weak spectral lines on optical wavelengths and is almost
featureless over $5200\,\AA$. After the maximum, the width of the IGE lines
in $3600-4400\,\AA$ increases with time, indicating the expansion of their
effective line-forming regions. This evolution supports the assumption of
uniform abundances of IGEs predicted by pure deflagration models. The strength
of IGE lines increases moderately in the $4400-5400\,\AA$ range during the
observed post-maximum epochs, while the Na\,I\,D line and the Ca\,II\,NIR
triplet display a continuous increase. Due to the lower excitation temperature
(see, e.g., \cite{NaID-exct-temp}) and the blue shift of the Na\,I\,D line
indicates a high abundance of Na in the outermost region of the ejecta. The
Si\,II\,$\lambda6355$ line is clearly visible during the early epochs, but
gradually weakens at later epochs. 

\begin{figure*}
    \centering
    \includegraphics[width=0.90\linewidth]{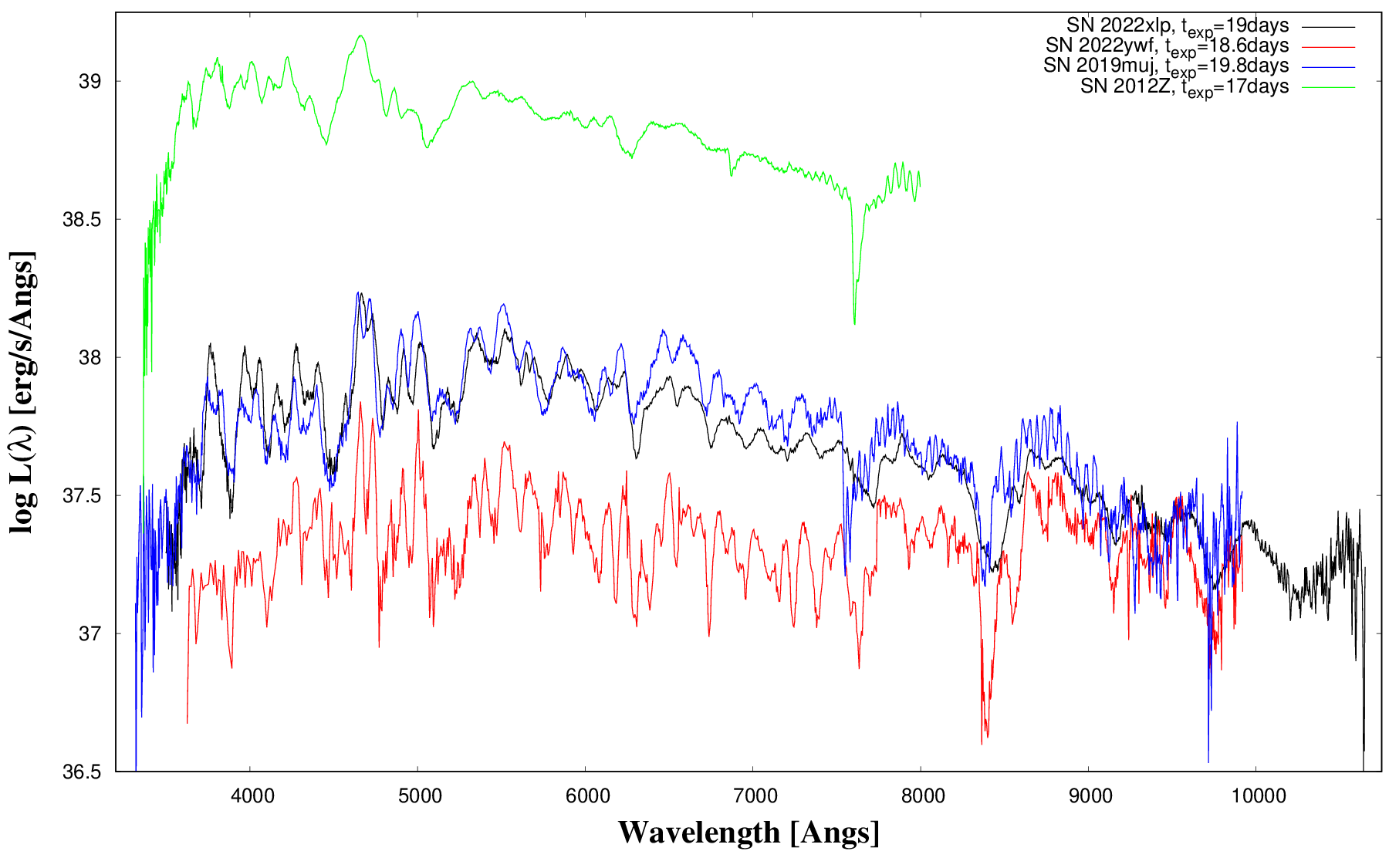}
    \caption{Spectra of SN 2022xlp, SN 2019muj \citep{SN2019muj}, SN2012Z
\citep{SN2012Z}, and SN 2022ywf (Barna et al. in prep.) measured 19 days
after the explosion. Because of the order of magnitude difference in luminosity
in the Iax subclass, we plot the logarithm of spectral luminosity density.
The strong similarity between SN 2022xlp and SN 2019muj can be amply observed.}
    \label{fig:Iax-spectra-comparison}
\end{figure*}

\begin{figure*}
    \centering
    \includegraphics[width=0.90\linewidth]{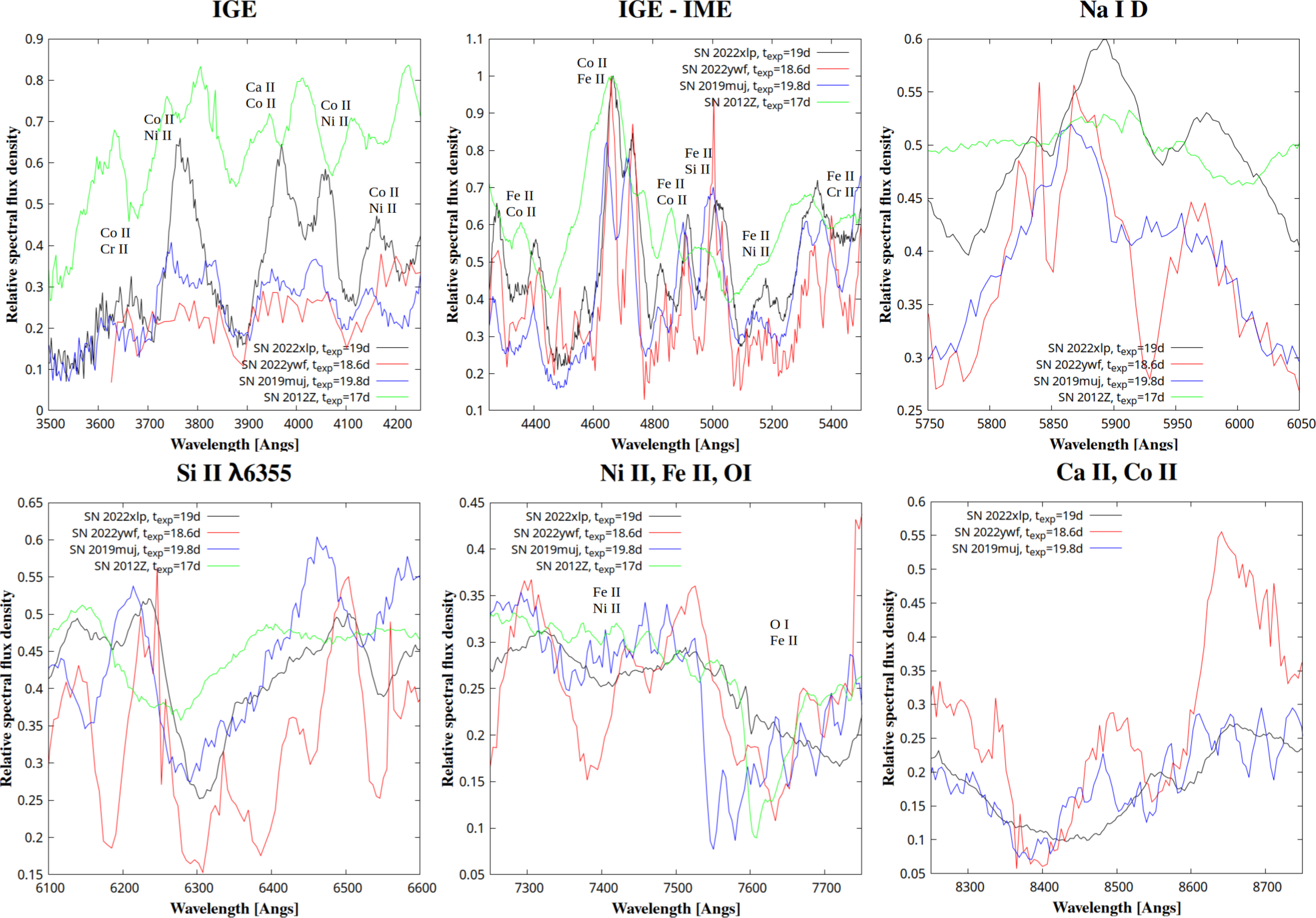}
    \caption{Normalized spectra of the Iax-type supernovae SN 2019muj \citep{SN2019muj},
the intermediate-luminous SN 2022xlp, bright SN 2012Z \citep{SN2012Z}, and
faint SN 2022ywf (Barna et al. in prep.) in the wavelength range of the characteristic
P-Cygni spectral lines.  Note: the S/N  of the spectra of SN 2012Z is too
low above $\sim 8000\,\AA$  and thus it is not plotted in the lower right
subplot.}
    \label{fig:Iax-spectra-comparison-lines}
\end{figure*}

We also compared the spectra of four SNe from different luminosity groups
of the Iax subclass at similar phases.
All spectra obtained approximately 19 days after the explosion, thus relatively
close to their peak luminosity, are plotted in Fig. \ref{fig:Iax-spectra-comparison}
(when the characteristic IME spectral features can still be identified).
There is a strong similarity between SN 2022xlp and SN 2019muj with respect
to their continuum and spectral lines. Figure \ref{fig:Iax-spectra-comparison}
shows that the photospheric temperature and, thus, the steepness of the continuum
increase with the brightness of the supernova at the same epoch relative
to the explosion. The characteristic spectral lines of the Iax SN sample
can be compared by normalization of the spectra, which are plotted in Fig.
\ref{fig:Iax-spectra-comparison-lines}. In the case of SN 2022xlp, the lines
of IGE elements, namely Co\,II, Ni\,II, Ca\,II, and Cr\,II are stronger in
the $3700-4750\,\AA$ range compared to SN 2019muj, which is a result of the
difference in the density profile due to the difference in epochs between
the two spectra. Above $4750\,\AA$, we observe lines of approximately the
same strength, such as Fe\,II, Co\,II, and Si\,II.
In the case of the fainter SN 2022ywf, the Na\,I\,D line is stronger than
in either of the other supernovae; this is also the case for the Fe\,II,
Ni\,II, and O\,I lines in the $7300-7700\,\AA$ range, as well as the Ca\,II
and Co\,II lines in the $8300-8700\,\AA$ range. The more luminous SN 2012Z
shows the same set of lines in the $3500-5400\AA$ wavelength range. The Si\,II\,$\lambda6355$
line is the weakest in this supernova sample, while the Na\,I\,D cannot be
identified. The prominent appearance of Na\,I\,D and Si\,II\,$\lambda6355$
lines in the spectra of the fainter SNe Iax suggest either a lower ejecta
temperature or an enhancement of the IMEs. In general, the appearance of
these spectral features indicates similar abundance profiles but a faster
evolution of the fainter SNe Iax.
These differences may possibly arise from the steepness of their density
profiles. An alternative explanation is that SNe Iax have similar temperature
profiles, but the abundance profiles are stratified; at the same time, after
the explosion, the photosphere is located at a different layer. However,
the pure deflagration hydrodynamic simulations do not support a stratified
abundance profile scenario. 

We performed a detailed spectral tomography analysis on the spectral time
series of SN 2022xlp to determine the physical and abundance structure of
the ejecta. In a co-moving frame, the photosphere moves inward with time
as the ejecta expands and the temperature decreases. Thus, deeper and deeper
ejecta regions gradually become part of the optically thin atmosphere and
the strongest line-forming region. By comparing spectra taken at different
epochs, we can determine the radial profiles of the physical properties and
abundances via the fittings with synthetic spectra calculated by a radiative
transfer code. This method is called spectral tomography. We used a Monte
Carlo-based 1D radiative transfer code {\tt TARDIS} \citep{TARDIS-main} to
synthesize the theoretical model spectra. The model parameters kept fixed
during the fitting are presented in Appendix \ref{appendix:TARDIS-settings}.
The self-consistent fitting of spectra time series with TARDIS-synthesized
spectra allowed us to reveal the SN ejecta's radial density and abundance
profile. The parameters of these profiles must be specified at a reference
time and kept constant for all epochs. This is because the slope of the density
profile and the abundances of the stable isotopes do not change over time;
in addition, the code calculates the expansion-caused dilution together with
the radioactive decay of unstable isotopes. By comparing the parameters determined
from spectral tomography with those modelled from the pure deflagration explosion
simulations, we can estimate the possible explosion process and the progenitor
object of the supernova.

\begin{figure*}
    \centering
    \includegraphics[width=1\linewidth]{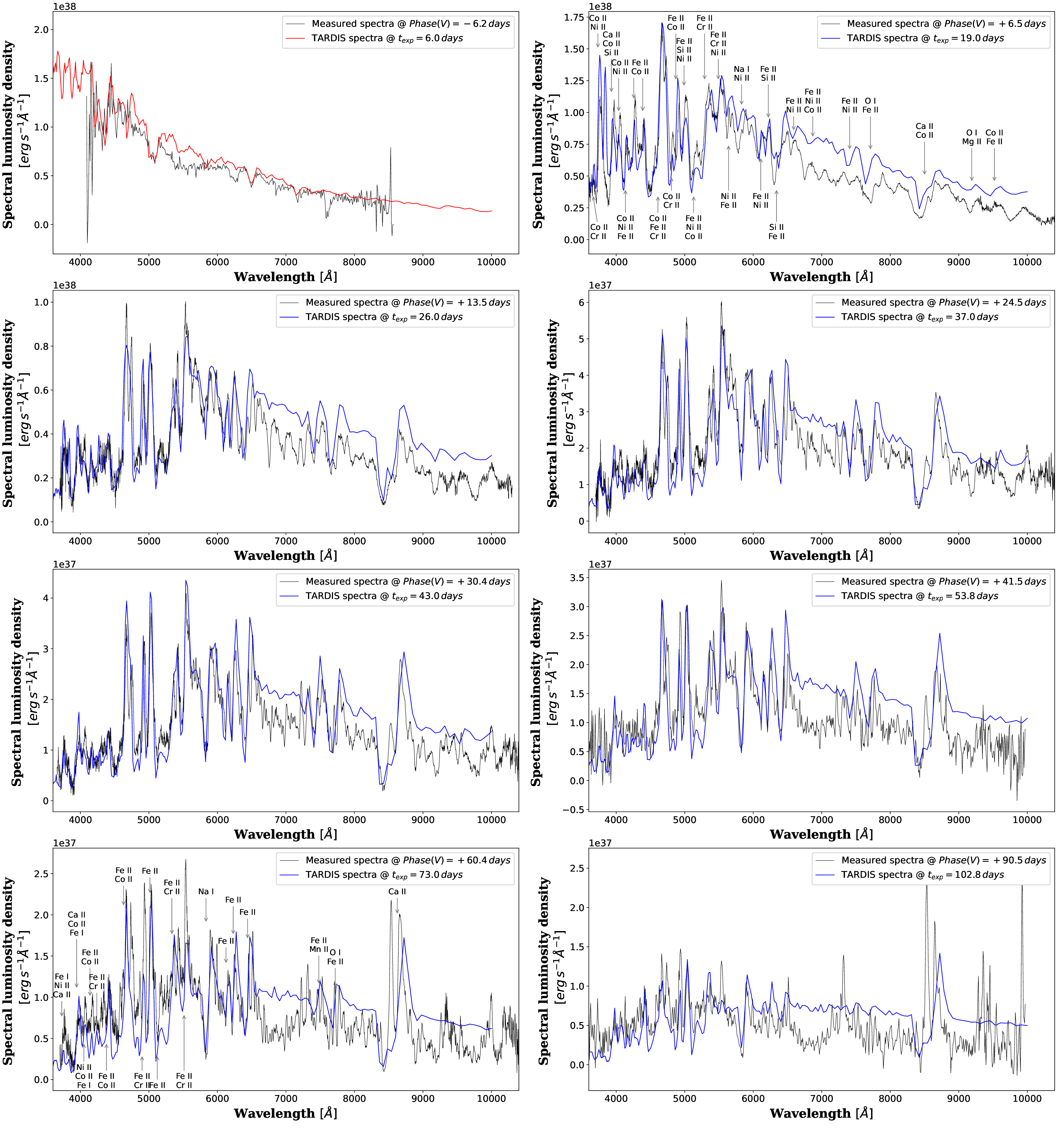}
    \caption{Spectra time series fittings of SN 2022xlp with TARDIS-synthesised
model spectra. The figure also shows the identified
    spectral lines at the two highlighted epochs: a post- and near maximum
and a late-phase spectrum. The phase relative to the V-band maximum and the
time since the explosion are also marked.}
    \label{fig:SN2022xlp-TARDIS-fits-multiple-texp}
\end{figure*}

\begin{figure*}
    \centering
    \includegraphics[width=1\linewidth]{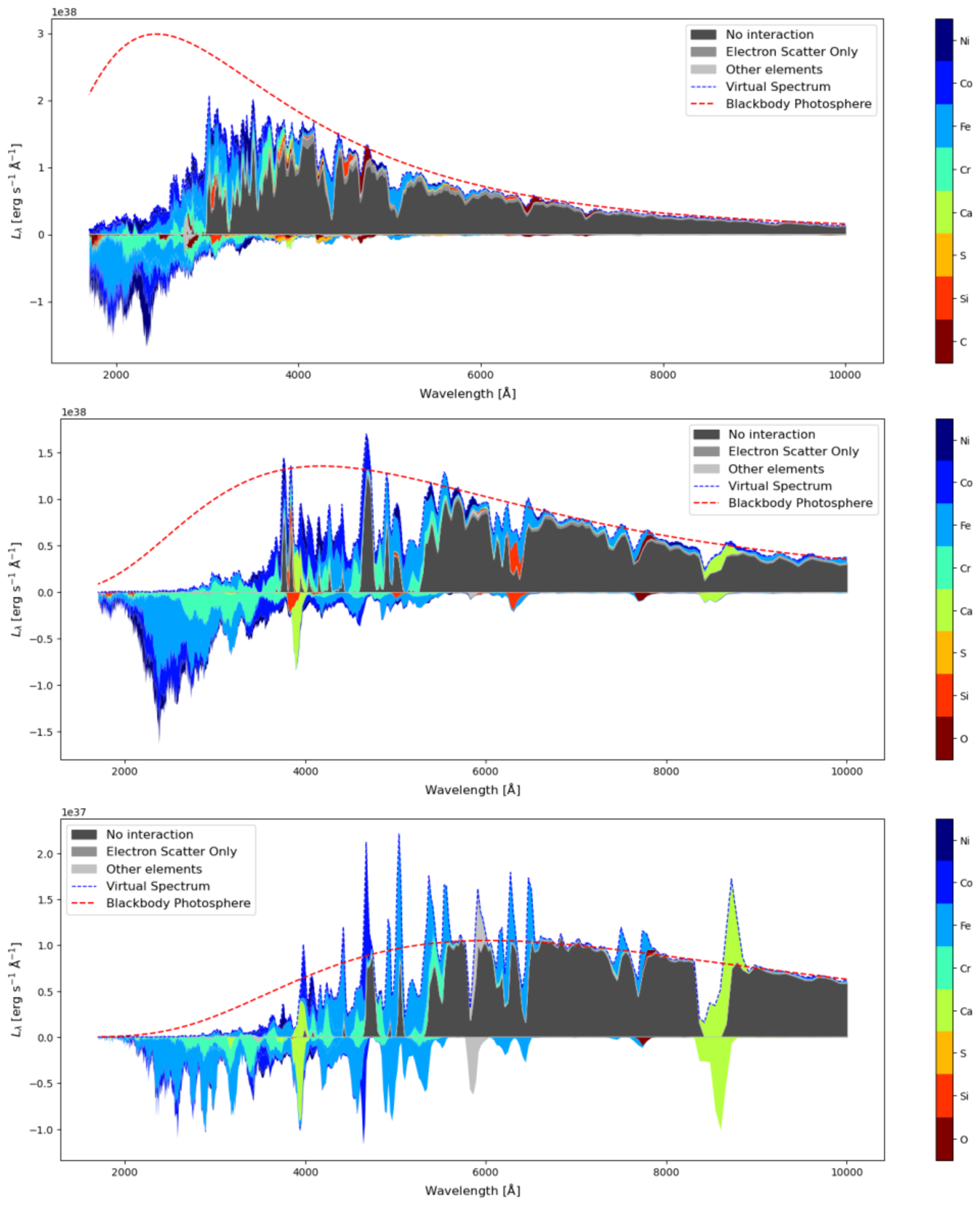}
    \caption{The spectral decomposition (SDEC) plots of SN 2022xlp at $t_\mathrm{exp}=6.0\,\mathrm{days}$
(top), $t_\mathrm{exp}=19.0\,\mathrm{days}$ (middle) and $t_\mathrm{exp}=73.0\,\mathrm{days}$
(bottom). The figures show the synthetic model spectrum and also the contribution
of various light-matter interactions and elements to the spectrum, indicating
how much they add or subtract from the spectral luminosity density at different
wavelengths. The red dashed line represents the Planck curve corresponding
to the photospheric temperature.}
    \label{fig:SN2022xlp-SDEC-series}
\end{figure*}

One of the most important sources of information  regarding the outermost
regions of the ejecta is the pre-maximum spectrum and its {\tt TARDIS} model
fitting, which can be seen in the upper left subplot of Fig. \ref{fig:SN2022xlp-TARDIS-fits-multiple-texp}.
It was observed $4.7\,\mathrm{days}$ and $6.2\,\mathrm{days}$ before the
maximum brightness of the B and V bands and $6\,\mathrm{days}$ after the
explosion date constrained in the spectral tomography analysis. The first
spectrum shows only a few relatively weak spectral lines; thus, it offers
the best opportunity to determine the characteristic value of continuum-related
physical quantities. The determined bolometric luminosity is $ L(t_\mathrm{exp}=6.0\,\mathrm{days})
= 1.58\times10^{8}\,L_{\odot}$, while the photospheric velocity is $v_\mathrm{phot}(t_\mathrm{exp}=6.0\,\mathrm{days})
= 5400\,\mathrm{km/s}$. The physical properties of the ejecta determined
by spectral tomography are summarized in Table \ref{tab:SN2022xlp-physical-parameters}
for all epochs. The strongest lines can be identified at shorter wavelengths
between $4100 - 5100\,\AA$, and belong to highly ionized IGE-s, mainly, Fe\,III,
Co\,III, and Ni\,II. There are some signs of IME-s, such as S\,III around
$4225\,\AA$, Si\,III at $4529\,\AA$, and C\,II around $4670\,\AA$, $6580\,\AA$
and $7170\,\AA$. The highly ionised ions and the steep, blue-dominated continuum
show high photospheric temperature. According to the fitted model spectra,
the photospheric temperature was $T_\mathrm{phot}(t_\mathrm{exp}=6.0\,\mathrm{days})
= 11870\,\mathrm{K}$. The Si\,II\,$\lambda6355$ line is clearly detectable.
The effects of the Thomson-scattering and the bound-bound transitions of
different ions can be visualized in a so-called spectral decomposition (SDEC)
plot, shown in Fig. \ref{fig:SN2022xlp-SDEC-series}. The SDEC plot of the
first epoch presents the dominance of Ni, Co, and Fe ions in the spectra,
as we would expect from the hydrodynamic deflagration simulations (see models
from, e.g., \cite{Iax-def-Lach} or  \cite{Iax-def-Fink}). Due to the lack
of very early-phase photometric data, we were unable to determine the explosion
time using the expanding fireball method, so we treated it as a fitting parameter
during the spectral fitting process to obtain this parameter. 

The next spectrum of the time series was measured $19.0\,\mathrm{days}$ after
the explosion, as well as $8.0\,\mathrm{days}$ and $6.5\,\mathrm{days}$ after
the B and V maximum, respectively. Compared with the spectrum of the previous
epoch, it shows stronger lines over the whole wavelength range. The most
prominent line-forming IGE ions are Co\,II, Ni\,II, Fe\,II, and Cr\,II; while
the Ca\,II, Si\,II és O\,I lines from IME also have significant features,
especially in the red part of the spectrum. The Ca\,II also has an effect
on the blue part and Mg\,II contributes to a spectral line above $9000\,\AA$.
The observed and fitted synthetic spectra are plotted in the top right panel
of Fig. \ref{fig:SN2022xlp-TARDIS-fits-multiple-texp}. With the decrease
in photospheric temperature, the Si\,II$\lambda$6355 line emerged more prominently.
According to the best-fit TARDIS model spectrum, the photospheric temperature
was $T_\mathrm{phot}(t_\mathrm{exp} = 19.0\,\mathrm{days}) = 6937\,K$, its
velocity $v_\mathrm{phot}(t_\mathrm{exp} = 19.0\,\mathrm{days}) = 4400\,\mathrm{km/s}$,
while the luminosity $L(t_{\mathrm{exp}} = 19.0\,\mathrm{days}) = 1.58\times10^{8}\,L_{\odot}$.
We also measured the photospheric velocity by fitting the Si\,II\,$\lambda
6355$ line, which is a widely used method for thermonuclear SNe. In case
of type Iax SNe, however, Si\,II\,$\lambda 6355$ is often blanketed by Fe\,II\,$\lambda
6456$ and by Co\,II lines, which may lead to the underestimation of $v_{phot}$
when applying a single Gaussian profile to fit the observed absorption feature
 \citet{SN2011ay-1}. Therefore, we also adopted a double Gaussian function
to account for the presence of any other lines next to Si\,II\,$\lambda 6355$.
This method provides $v_\mathrm{phot,Gauss}(t_\mathrm{exp} = 19.0\,\mathrm{days})
= 2570\,\mathrm{km/s}$, which is still a significantly lower value than that
of the spectral tomography. Since radiative transfer modeling offers a more
physical approach and, thus, more accurate velocity estimation is expected,
we accept and use $v_\mathrm{phot}(t_\mathrm{exp} = 19.0\,\mathrm{days})
= 4400\,\mathrm{km/s}$ constrained by spectral synthesis hereafter. The SDEC
plot of this near-maximum epoch can be seen in Fig. \ref{fig:SN2022xlp-SDEC-series},
which still shows significant Ni, Co, and Fe spectral features. 

The fit of the next four spectra in the time series can be seen in the two
middle rows of Fig. \ref{fig:SN2022xlp-TARDIS-fits-multiple-texp}. The phases
and the time since the explosion are listed in Table \ref{tab:SN2022xlp-spectra-data},
while the determined time-dependent physical parameters, such as the photospheric
velocity and temperature, as well as the bolometric luminosity, are summarized
in Table \ref{tab:SN2022xlp-physical-parameters}.  The synthetic spectrum
agrees well with the observed spectrum below $6600\,\AA$. Above this wavelength,
the model continuum overestimates the observed spectrum, which is caused
either by the uncertainty of photospheric properties or the presence of unidentified
absorption features. The spectral luminosity density decreases monotonically
after the maximum, with an initially faster rate of decline. Similar conclusions
can be drawn regarding the photospheric temperature and velocity. In the
shorter wavelength part of the spectrum, below $5500\,\AA$, the main line-forming
ions from the IGE elements are Co\,II, Fe\,II, and Ni\,II, while from the
IME elements are Cr\,II, Ca\,II, and Si\,II. The strength of the Ni feature
decreases with time due to the radioactive decay of $^{56}$Ni. On the red
side of the spectrum, above $5500\,\AA$, the line-forming ions are Na\,I,
Fe\,II, Cr\,II, Si\,II, Ca\,II, and Mg\,II. A detectable line is also produced
by O\,I around $\sim 7750\,\AA$, as well as a significant line triplet by
Ca\,II near $\sim 8830\,\AA$, which strengthens over time. As time progresses
after the explosion, the spectrum becomes increasingly dominated by Fe\,II
ion lines, and the Na\,I\,D line also becomes more prominent due to the temperature
decrease in the atmospheric layers. Some spectral lines in the $6650 - 7320\,\AA$
range, as well as those near $\sim 9760\,\AA$, could not be fitted. The presence
of these lines likely requires modification of the abundance profile adopted
from deflagration models used for comparison, requiring adjustments to the
abundances of certain isotopes (introduced at the end of this section). 

We used two of the last four observed spectra of SN 2022xlp for our spectral
tomography, since the late-phase evolution of thermonuclear SNe is slower
than in the early phase. These late-time spectra were obtained at $t_\mathrm{exp}
= 73.0\,\mathrm{days}$ and $t_\mathrm{exp} = 102.8\,\mathrm{days}$ after
the explosion ($60.4\,\mathrm{days}$ and $90.5\,\mathrm{days}$ after the
V-band maximum). The observed and fitted model spectra are plotted in Fig.
\ref{fig:SN2022xlp-TARDIS-fits-multiple-texp}. By this time, the spectrum
is dominated by Fe\,I and Fe\,II lines throughout the optical regime. The
only noticeable contributions of IMEs are the Na\,I\,D line, the Ca\,II NIR
triplet, and the O\,I $\lambda7770$ feature. At shorter wavelengths, there
are weak P~Cygni profiles of Ca\,II, Cr\,II, and Co\,II ions. The contributions
of Cr\,II and O\,I ions nearly disappear 102.8 days after the explosion.
Beyond $6600\,\AA,$ the continuum flux of the synthetic spectrum is higher
than that of the observed spectrum and fitting the observed spectral features
between $6600-7360\,\AA$, $7780-8310\,\AA$ and above $8910\,\AA$ was not
feasible with our TARDIS model. A possible explanation of this discrepancy
in the $7000-7500\,\AA$ region is the rise of forbidden emission lines in
the late phase spectrum, described in \citet{late-time-Iax-spectra}. Forbidden
emissions are significant at 102.8 days (prior to this, the growing forbidden
lines are difficult to distinguish from the previously existing P Cygni features),
where we can see some narrow and strong emission lines in the red part of
the spectrum. According to the discussion of \citet{late-time-Iax-spectra},
the Fe\,II\,$\lambda7155$, Ca\,II\,$\lambda7291$, Ca\,II\,$\lambda7324$,
and Ni\,II\,$\lambda7378$ lines are the strongest forbidden emission lines.
The appearance of these emission features is made possible by the significant
dilution of the outer layers of the supernova ejecta; however, the TARDIS
code is unable to synthesize these lines. Thus, we simply omitted these emission
features from our fitting process. However, based on TARDIS fits, the Fe\,II\,$\lambda7465$
line appears, which is not a forbidden emission line at least at 73.0 days
after the explosion. The photospheric velocity has also decreased to $v_\mathrm{phot}(t_\mathrm{exp}
= 73.0\,\mathrm{days}) = 800\,\mathrm{km\,s^{-1}}$ and further decreased
to $v_\mathrm{phot}(t_\mathrm{exp} = 102.8\,\mathrm{days}) = 565\,\mathrm{km\,s^{-1}}$,
while the photospheric temperature has also changed to $T_\mathrm{phot}(t_\mathrm{exp}
= 73.0\,\mathrm{days}) = 4803\,K$ and $T_\mathrm{phot}(t_\mathrm{exp} = 102.8\,\mathrm{days})
= 4475\,K$. The bolometric luminosity has reached $L(t_\mathrm{exp} = 73.0\,\mathrm{days})
= 2.34\times10^{7}\,L_{\odot}$ and $L(t_\mathrm{exp} = 102.8\,\mathrm{days})
= 1.82\times10^{7}\,L_{\odot}$. The SDEC plot of this epoch can be seen in
Fig. \ref{fig:SN2022xlp-SDEC-series}, which shows exclusively Fe-dominated
spectra and the significant Na\,I\,D and Ca\,II NIR triplet features. 

\begin{figure}
    \centering
    \includegraphics[width=1\linewidth]{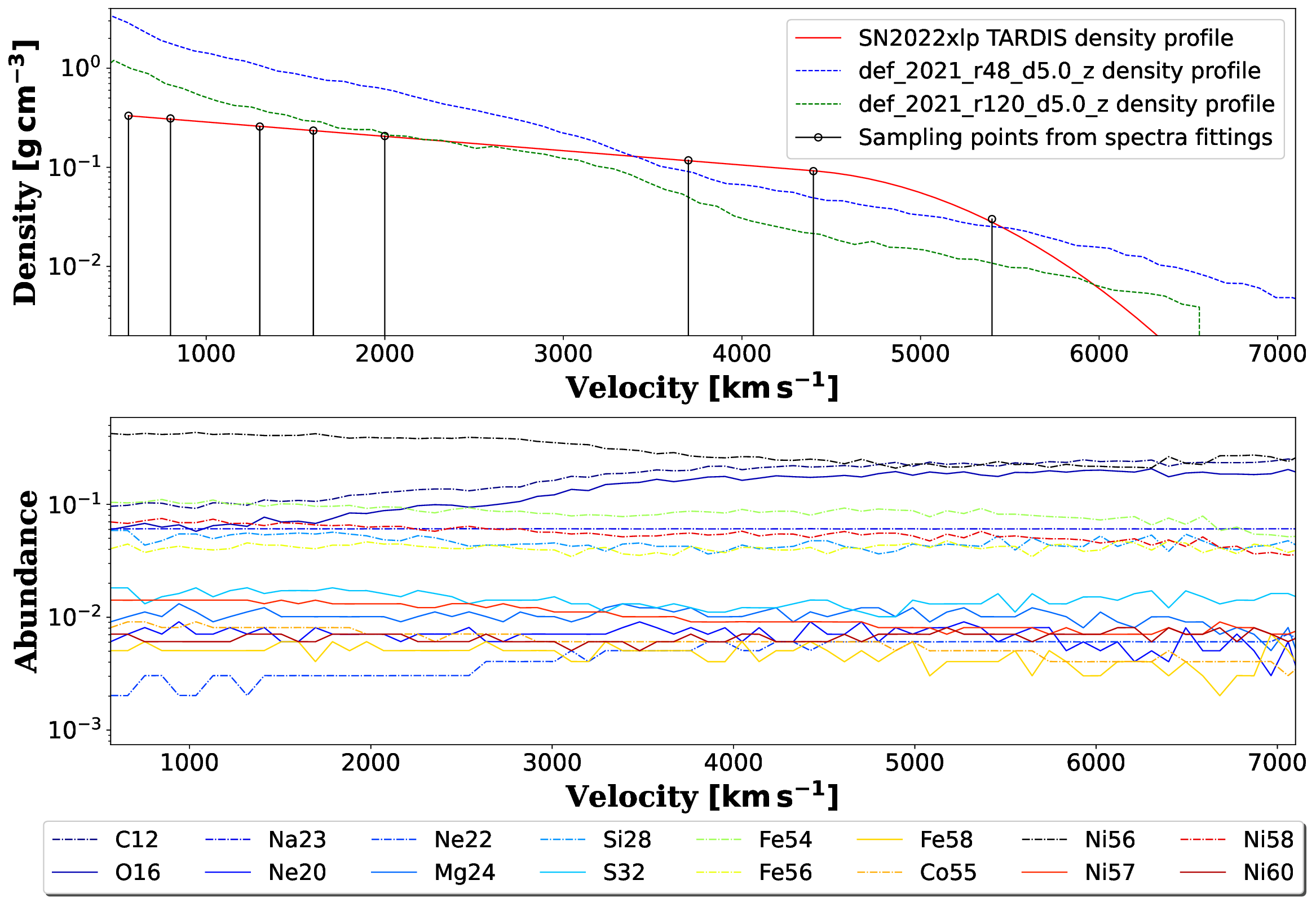}
    \caption{Density and abundance profiles obtained from the tomographic
analysis of SN 2022xlp. Vertical lines on the density profile indicate its
sampling points. The blue and green dashed lines show the density profiles
of the {\tt def\_2021\_r48\_d5.0\_z} and {\tt def\_2021\_r120\_d5.0\_z} models.
The abundance profile is identical to that of the {\tt def\_2021\_r48\_d5.0\_z}
deflagration model, except for an increased Na abundance and a corresponding
reduction in O abundance.}
    \label{fig:SN2022xlp-abudenplot}
\end{figure}

\begin{figure}
    \centering
    \includegraphics[width=1\linewidth]{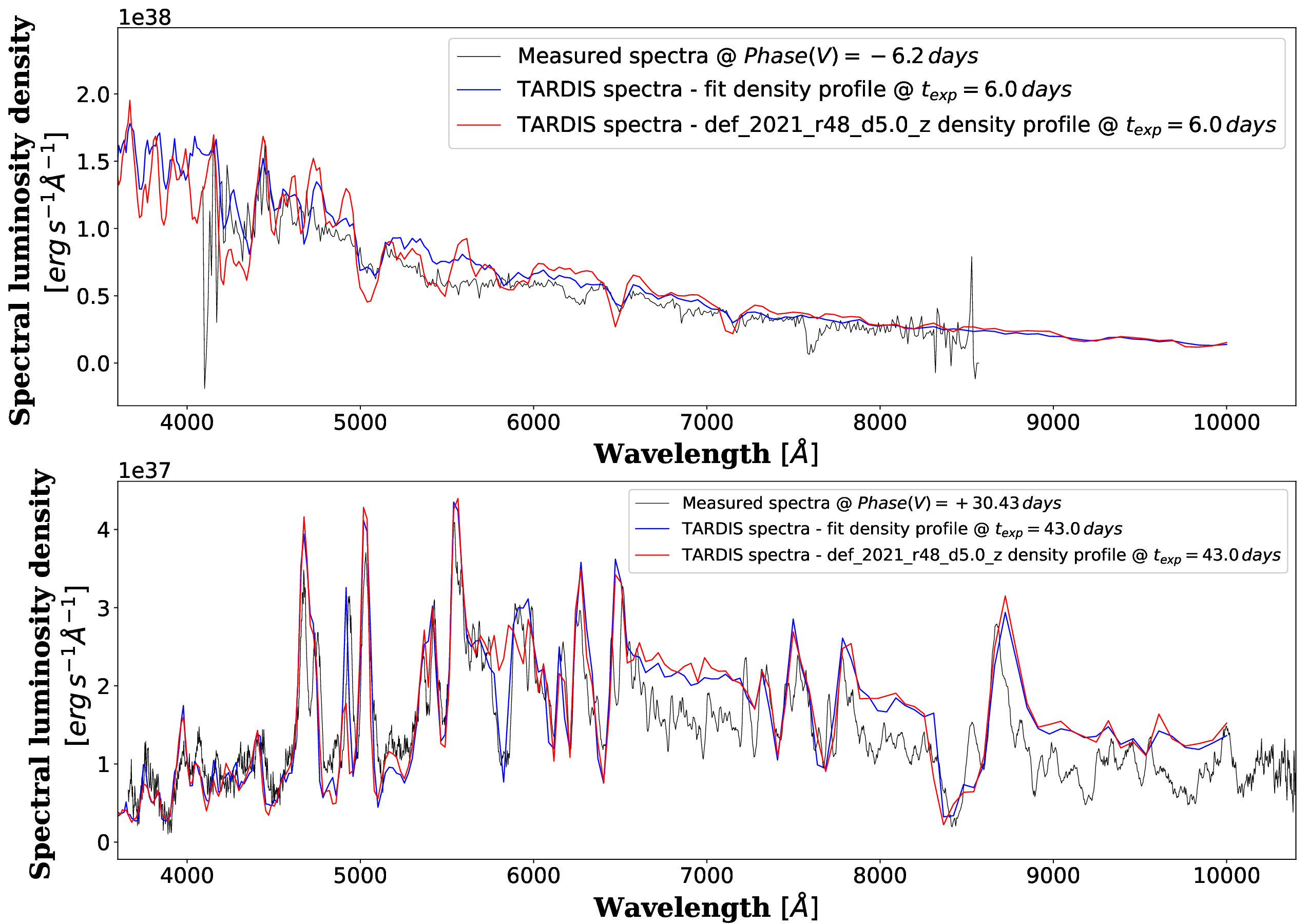}
    \caption{Comparison of the model spectra of SN 2022xlp at different epochs.
The model spectra calculated with the fitted density and Na-corrected abundance
profile are shown with blue curves. In contrast, the spectra computed using
exactly the adopted density and abundance profiles of the {\tt def\_2021\_r48\_d5.0\_z}
model are shown with red curves. Apart from the density and abundance profiles,
the values of the other parameters are identical at each epoch.}
    \label{fig:SN2022xlp-density-profile-comparsion}
\end{figure}

From the various pure deflagration models of \citet{Iax-def-Fink} and \citet{Iax-def-Lach},
we selected those  exhibiting a peak luminosity and bolometric LC roughly
similar to those of SN 2022xlp. The most similar models are {\tt def\_2021\_r48\_d5.0\_z}
and {\tt def\_2021\_r120\_d5.0\_z} from \citet{Iax-def-Lach} with peak bolometric
absolute magnitudes of $M_{\mathrm{bol, peak}}(\mathrm{def\_2021\_r48\_d5.0\_z})
= -15.83\,\mathrm{mag}$ and $M_{\mathrm{bol, peak}} (\mathrm{def\_2021\_r120\_d5.0\_z})
=-15.38\,\mathrm{mag}$. We adopted the abundance profile of model {\tt def\_2021\_r48\_d5.0\_z}
for the spectral synthesis with TARDIS, but allowed for minor modifications
to further improve the fit. Regarding the density profile, we applied a custom
model based on \citet{SN2019muj} or \citet{Iax-few-param-family}, which can
be expressed as:
\begin{equation}
    \rho(v) = \begin{cases}
    \rho_{0} \times e^{-\frac{v}{v_{0}}} & \mbox{if} \hspace{0.1cm} v \le
v_{lim}, \\
    \rho_{0} \times e^{-\frac{v}{v_{0}}} \times 8^{-\frac{8(v-v_{lim})^2}{v_{lim}^2}}
& \mbox{if} \hspace{0.1cm} v > v_{lim},
    \end{cases}
\label{eq:Deflagration-density-profile}
\end{equation}
where $\rho_{0}$ is the core density at $100\,\mathrm{s}$ after the explosion,
$v_\mathrm{0}$ describes the decrease in density outward the ejecta. 
The derived density profile, along with the adopted {\tt def\_2021\_r48\_d5.0\_z}
and minor-modified deflagration abundance profile, is shown in Fig. \ref{fig:SN2022xlp-abudenplot}.
The density profile of the {\tt def\_2021\_r48\_d5.0\_z} model produces overly
strong spectral lines at early epochs, as shown in Fig. \ref{fig:SN2022xlp-density-profile-comparsion},
because it predicts excessively high densities in the layers above the photosphere.
To improve the fit of synthetic spectra, we constrained a density function
described by Eq. \ref{eq:Deflagration-density-profile} instead, where the
parameters are $\rho_0 = 0.4\,\mathrm{g\,cm^{-3}}$, while $v_0 = 3000\,\mathrm{km\,s^{-1}}$
(summarized in Table \ref{tab:SN2022xlp-physical-parameters}). Our fitted
density profile introduces a steep cutoff of $v_\mathrm{lim} = 4400\,\mathrm{km\,s^{-1}}$,
beyond which the density declines more rapidly, resulting in a proper fit
for the spectral lines. 
After the maximum, the spectra become less sensitive to the density profile,
as is seen in Fig. \ref{fig:SN2022xlp-density-profile-comparsion}. 
The slope of the constrained density profile differs from the density profiles
estimated by the deflagration models {\tt def\_2021\_r48\_d5.0z\_} and {\tt
def\_2021\_r120\_d5.0\_z}. The slightly slower decline leads to more material
being present before the cutoff point, and significantly less in the outermost
regions of the ejecta beyond the cutoff.  
The synthetic spectra closer to the maximum brightness are more sensitive
to the abundance profile, while the later spectra are dominated by iron lines
and insensitive to the IMEs. The adopted abundance profile of the {\tt def\_2021\_r48\_d5.0\_z}
deflagration model results in a very good fit to the series of spectra, except
for the Na\,I\,D line. This line could be fit only by significantly increasing
the Na abundance to $0.06$ in all layers, while correspondingly decreasing
the O abundance, which has a relatively weaker effect. The spectra calculated
with the density and abundance profile of the {\tt def\_2021\_r48\_d5.0\_z}
model are shown with our fitted spectra in Fig. \ref{fig:SN2022xlp-density-profile-comparsion}.
Here, we can see that the Na\,I\,D line cannot be fit with the abundance
profile of the {\tt def\_2021\_r48\_d5.0\_z} model.

\begin{table*}
\label{tab:SN2022xlp-physical-parameters}
\caption{Physical properties of the SN 2022xlp ejecta constrained in fitting
the spectral series with TARDIS. The parameters of the density profile are
listed in the top line (see in Eq. \ref{eq:Deflagration-density-profile}).
Here, $\rho_\mathrm{0}$ is the central density of the model ejecta 100~sec
after the explosion.}
\centering
\begin{tabular}{cccccc}
\hline \hline
\multicolumn{2}{c}{$\rho_{0}\,[g\,cm^{-3}]$} &
\multicolumn{2}{c}{$v_{0}\,[km\,s^{-1}]$}    &
\multicolumn{2}{c}{$v_{lim}\,[km\,s^{-1}]$} \\
\hline
\multicolumn{2}{c}{0.4}  &
\multicolumn{2}{c}{3000} &
\multicolumn{2}{c}{4400} \\
\hline \hline
MJD                        &
Phase (V)                  &
$t_{exp}\,[\mathrm{days}]$ &
$v_{phot}\,[km\,s^{-1}]$   &
$T_{phot}\,[K]$            &
$L\,[10^{7} L_{\odot}]$    \\
\hline
59867.83 & -6.2   & 6.0   & 5400 & 11870 & 15.8 \\
59880.52 & +6.5   & 19.0  & 4400 & 6937  & 15.8 \\
59887.53 & +13.5  & 26.0  & 3700 & 5686  & 10.6 \\
59898.53 & +24.5  & 37.0  & 2000 & 5480  & 5.6  \\
59904.49 & +30.4  & 43.0  & 1600 & 5312  & 4.5  \\
59915.59 & +41.5  & 53.7  & 1300 & 4935  & 3.6  \\
59934.50 & +60.4  & 73.0  & 800  & 4803  & 2.3  \\
59964.59 & +90.5  & 102.8 & 565  & 4475  & 1.8  \\
\hline \hline
\end{tabular}
\end{table*}

\FloatBarrier
\clearpage

\section{Summary and conclusions}
\label{section:summary}
We present our multicolor photometric and spectroscopic observations of SN
2022xlp, which is the second intermediate-luminosity type Iax SN with detailed
follow-up, as its V-band LC peaks at $M_{max}(V) = -16.04 \pm 0.25\,\mathrm{mag}$.
The data set starts at $6.0\,\mathrm{days}$ and covers $\sim 73.0\,\mathrm{days}$
after the explosion. 
Based on observing characteristics, such as LC properties, color evolution,
and spectral features, SN 2022xlp is shown to be very similar to SN 2019muj,
 bridging the luminosity gap of the diverse Iax class. The change in color-indexes
is $\sim1.5$ mag between $-8.0$ and $20.0$ days, indicating a rapidly decreasing
photospheric temperature. After $20\,\mathrm{days}$ the V-band maximum,
the B-V and r-i colors are almost constant, while g-r and g-i show a slow
decline. By comparing the color evolution of various luminous SNe Iax, we
can conclude that the amplitude of the color change is greater in the case
of brighter Iax supernovae.

We performed a semi-bolometric LC fitting based on the Arnett model to set
constraints on the general parameters of the ejecta. The results of the modeling
are $M_\mathrm{Ni} = 0.012 \pm 0.009\,M_{\odot}$ for the nickel mass, $M_\mathrm{ej}
= 0.142 \pm 0.015\,M_{\odot}$ for the ejecta mass, and $E_\mathrm{kin} =
(2.066\pm 0.236) \times 10^{49}\,\mathrm{erg}$ for the kinetic energy. Based
on the synthetic LC fitting, the Ni mass, ejecta mass, explosion energy,
and effective diffusion time are similar to those of SN 2019muj. According
to the bolometric LC analysis of SNe 2019muj and 2022xlp, the characteristic
mass parameters of the intermediate luminous SNe Iax are $M_\mathrm{Ni}\sim
0.02-0.03\,M_{\odot}$ and $M_\mathrm{ej} \sim 0.09-0.17\,M_{\odot}$.
We performed a comprehensive spectroscopic analysis that included a qualitative
comparison of the spectra of different SNe Iax with different peak luminosities,
a spectral line evolution study, and detailed spectral tomography. The main
results are listed below. 
\begin{enumerate}
    \item Based on the analysis of the evolution of spectral features: \begin{itemize}
        \item After the maximum, the widths of the IGE spectral lines at
the $3600-4400\,\AA$ wavelength range indicate a widening of the line-forming
regions of IGEs, along with quasi-constant abundances. This result shows
good agreement with pure deflagration hydrodynamic simulation results.
        \item The increasing strength of the Na\,I\,D lines indicates a high
Na abundance ($X(Na) \simeq 0.05$) in the outermost region of the ejecta,
compared to the predictions of pure deflagration models.
    \end{itemize}
    \item Based on the comparison of the spectra of different SNe Iax at
the same time since explosion:
    \begin{itemize}
        \item There is a strong similarity between SN 2022xlp and SN 2019muj
        \item The prominent appearance of Na\,I\,D and Si\,II\,$\lambda6355$
lines in the spectra of fainter SNe Iax suggest a lower photospheric temperature
profile in the case of fainter Iax supernovae. These spectral features also
indicate similar abundance profiles.
        \item The fainter SNe Iax evolve faster. Their temperature profile
decreases faster or the decline starts at lower temperatures. The difference
in the evolution rate can possibly originate from the difference in the density
profile.
    \end{itemize}
    \item Based on the spectral tomography:
    \begin{itemize}
        \item The physical quantities characteristic of the ejecta, the luminosity,
the photospheric temperature, and the velocity could be determined by fitting
the observed spectra time series with the {\tt TARDIS} synthesized spectra.
These results are summarized in Table \ref{tab:SN2022xlp-physical-parameters}.
        \item We present the line identification results at the most important
epochs, the first post-maximum spectra, and the latest spectra (see Fig.
\ref{fig:SN2022xlp-TARDIS-fits-multiple-texp}, and Fig. \ref{fig:SN2022xlp-SDEC-series})
        \item The density profile could be determined and compared with the
results of the most similar pure deflagration hydrodynamic simulations (see
 Fig. \ref{fig:SN2022xlp-abudenplot}). The most important result is that
the determined density profile differs in its rate of decline from the density
profiles calculated by the deflagration hydrodynamic simulations {\tt def\_2021\_r48\_d5.0\_z}
and {\tt def\_2021\_r120\_d5.0\_z}. 
        Our fitted density profile introduces a steep cutoff resulting significantly
lower densities in the outermost regions to prevent the presence of excessively
strong lines at the early epochs.
        \item The adopted abundance profile of the {\tt def\_2021\_r48\_d5.0\_z}
deflagration model results in a very good fit to the series of spectra, except
for the Na\,I\,D line. This line can only be well fitted by significantly
increasing the Na abundance in all layers, while correspondingly decreasing
the O abundance, which has a relatively smaller effect.
    \end{itemize}
\end{enumerate}
In summary, SN 2022xlp shows a very strong similarity to the first, well-analyzed
intermediate-luminous Iax supernova, SN 2019muj. SN 2022xlp supports the
assumption of the luminosity-velocity relation bridging the previously existing
luminosity gap together with SN 2019muj. For a more detailed recognition
of the intermediate-luminous range of SNe Iax, further analyses and even
more objects would be required.

\section*{Data availability}
All spectra are available at the WISeREP \citep{WISeREP} online supernova
database: \url{https://www.wiserep.org/object/21831}.

\begin{acknowledgements}
\label{section:acknow}
This work was supported by the professional funding from the University Research
Scholarship Programme (project code: EKÖP-24-3-SZTE-484) of the Ministry
of Culture and Innovation, financed by the National Research, Development,
and Innovation Fund (NKFIH), Hungary. The deployment and operation of the
BRC80 telescope was supported by the GINOP 2.3.2-15-2016-00033 project of
the NKFIH and the Hungarian Government, funded by the European Union. JV
is supported by the NKFIH-OTKA grant K142534. 

The UCSC team is supported in part by NASA grant 80NSSC20K0953, NSF grant
AST--1815935, and by a fellowship from the David and Lucile Packard Foundation
to R.J.F.

A major upgrade of the Kast spectrograph on the Shane 3 m telescope at the
Lick Observatory was made possible through generous
gifts from the Heising–Simons Foundation as well as William and
Marina Kast. Research at the Lick Observatory is partially supported
by a generous gift from Google.

Supernova research at Rutgers University is supported in part by US NSF award
AST-2407567 to S.W.J.

IRAF is distributed by NOAO, which is operated by AURA, Inc.,
under cooperative agreement with the NSF.

L.G. acknowledges financial support from AGAUR, CSIC, MCIN and AEI 10.13039/501100011033
under projects PID2023-151307NB-I00, PIE 20215AT016, CEX2020-001058-M, and
2021-SGR-01270.

This project has received funding from the HUN-REN Hungarian Research Network.
I. B. B. acknowledge the financial support of the Hungarian National Research,
Development and Innovation Office – NKFIH Grants K-147131 and K-138962.
 \end{acknowledgements}
\newpage
\clearpage

\bibliographystyle{aa}
\bibliography{example} 

\begin{thebibliography}{52}
\expandafter\ifx\csname natexlab\endcsname\relax\def\natexlab#1{#1}\fi

\bibitem[{{Arnett}(1982)}]{Arnett-1}
{Arnett}, W.~D. 1982, \apj, 253, 785

\bibitem[{{Barna} {et~al.}(2021){Barna}, {Szalai}, {Jha}, {Camacho-Neves}, {Kwok}, {Foley}, {Kilpatrick}, {Coulter}, {Dimitriadis}, {Rest}, {Rojas-Bravo}, {Siebert}, {Brown}, {Burke}, {Padilla Gonzalez}, {Hiramatsu}, {Howell}, {McCully}, {Pellegrino}, {Dobson}, {Smartt}, {Swift}, {Stacey}, {Rahman}, {Sand}, {Andrews}, {Wyatt}, {Hsiao}, {Anderson}, {Chen}, {Della Valle}, {Galbany}, {Gromadzki}, {Inserra}, {Lyman}, {Magee}, {Maguire}, {M{\"u}ller-Bravo}, {Nicholl}, {Srivastav}, \& {Williams}}]{SN2019muj}
{Barna}, B., {Szalai}, T., {Jha}, S.~W., {et~al.} 2021, \mnras, 501, 1078

\bibitem[{{Barna} {et~al.}(2018){Barna}, {Szalai}, {Kerzendorf}, {Kromer}, {Sim}, {Magee}, \& {Leibundgut}}]{Iax-few-param-family}
{Barna}, B., {Szalai}, T., {Kerzendorf}, W.~E., {et~al.} 2018, \mnras, 480, 3609

\bibitem[{{Branch} {et~al.}(2006){Branch}, {Dang}, {Hall}, {Ketchum}, {Melakayil}, {Parrent}, {Troxel}, {Casebeer}, {Jeffery}, \& {Baron}}]{Branch-normal-Ia}
{Branch}, D., {Dang}, L.~C., {Hall}, N., {et~al.} 2006, \pasp, 118, 560

\bibitem[{{Burrows} {et~al.}(2005){Burrows}, {Hill}, {Nousek}, {Kennea}, {Wells}, {Osborne}, {Abbey}, {Beardmore}, {Mukerjee}, {Short}, {Chincarini}, {Campana}, {Citterio}, {Moretti}, {Pagani}, {Tagliaferri}, {Giommi}, {Capalbi}, {Tamburelli}, {Angelini}, {Cusumano}, {Br{\"a}uninger}, {Burkert}, \& {Hartner}}]{Swift-1}
{Burrows}, D.~N., {Hill}, J.~E., {Nousek}, J.~A., {et~al.} 2005, \ssr, 120, 165

\bibitem[{{Chatzopoulos} {et~al.}(2012){Chatzopoulos}, {Wheeler}, \& {Vinko}}]{Arnett-MINIM-Vinko}
{Chatzopoulos}, E., {Wheeler}, J.~C., \& {Vinko}, J. 2012, \apj, 746, 121

\bibitem[{{Chatzopoulos} {et~al.}(2013){Chatzopoulos}, {Wheeler}, {Vinko}, {Horvath}, \& {Nagy}}]{MINIM}
{Chatzopoulos}, E., {Wheeler}, J.~C., {Vinko}, J., {Horvath}, Z.~L., \& {Nagy}, A. 2013, \apj, 773, 76

\bibitem[{{Coulter} {et~al.}(2023){Coulter}, {Jones}, {McGill}, {Foley}, {Aleo}, {Bustamante-Rosell}, {Chatterjee}, {Davis}, {Dickinson}, {Engel}, {Gagliano}, {Jacobson-Gal{\'a}n}, {Kilpatrick}, {Kutcka}, {Le Saux}, {Malanchev}, {Pan}, {Qui{\~n}onez}, {Rojas-Bravo}, {Siebert}, {Taggart}, {Tinyanont}, \& {Wang}}]{Coulter23}
{Coulter}, D.~A., {Jones}, D.~O., {McGill}, P., {et~al.} 2023, \pasp, 135, 064501

\bibitem[{Coulter {et~al.}(2022)Coulter, Jones, McGill, Foley, Aleo, Bustamante-Rosell, Chatterjee, Davis, Engel, Gagliano, Jacobson-Galán, Kilpatrick, Pan, Rojas-Bravo, Siebert, Taggart, Tinyanont, \& Wang}]{Coulter22}
Coulter, D.~A., Jones, D.~O., McGill, P., {et~al.} 2022, {YSE-PZ: An Open-source Target and Observation Management System}, {D. A. Coulter acknowledges support from the National Science Foundation Graduate Research Fellowship under Grant DGE1339067.}

\bibitem[{{Fink} {et~al.}(2014){Fink}, {Kromer}, {Seitenzahl}, {Ciaraldi-Schoolmann}, {R{\"o}pke}, {Sim}, {Pakmor}, {Ruiter}, \& {Hillebrandt}}]{Iax-def-Fink}
{Fink}, M., {Kromer}, M., {Seitenzahl}, I.~R., {et~al.} 2014, \mnras, 438, 1762

\bibitem[{{Firth} {et~al.}(2015){Firth}, {Sullivan}, {Gal-Yam}, {Howell}, {Maguire}, {Nugent}, {Piro}, {Baltay}, {Feindt}, {Hadjiyksta}, {McKinnon}, {Ofek}, {Rabinowitz}, \& {Walker}}]{Ia-LC-rise}
{Firth}, R.~E., {Sullivan}, M., {Gal-Yam}, A., {et~al.} 2015, \mnras, 446, 3895

\bibitem[{{Foley} {et~al.}(2009){Foley}, {Chornock}, {Filippenko}, {Ganeshalingam}, {Kirshner}, {Li}, {Cenko}, {Challis}, {Friedman}, {Modjaz}, {Silverman}, \& {Wood-Vasey}}]{SN2008ha}
{Foley}, R.~J., {Chornock}, R., {Filippenko}, A.~V., {et~al.} 2009, \aj, 138, 376

\bibitem[{{Foley} {et~al.}(2016{\natexlab{a}}){Foley}, {Jha}, {Pan}, {Zheng}, {Bildsten}, {Filippenko}, \& {Kasen}}]{Iax-spec-late}
{Foley}, R.~J., {Jha}, S.~W., {Pan}, Y.-C., {et~al.} 2016{\natexlab{a}}, \mnras, 461, 433

\bibitem[{{Foley} {et~al.}(2016{\natexlab{b}}){Foley}, {Jha}, {Pan}, {Zheng}, {Bildsten}, {Filippenko}, \& {Kasen}}]{late-time-Iax-spectra}
{Foley}, R.~J., {Jha}, S.~W., {Pan}, Y.-C., {et~al.} 2016{\natexlab{b}}, \mnras, 461, 433

\bibitem[{{Foley} {et~al.}(2016{\natexlab{c}}){Foley}, {Pan}, {Brown}, {Filippenko}, {Fox}, {Hillebrandt}, {Kirshner}, {Marion}, {Milne}, {Parrent}, {Pignata}, \& {Stritzinger}}]{Iax-Remnant}
{Foley}, R.~J., {Pan}, Y.-C., {Brown}, P., {et~al.} 2016{\natexlab{c}}, \mnras, 461, 1308

\bibitem[{{Foley} {et~al.}(2003){Foley}, {Papenkova}, {Swift}, {Filippenko}, {Li}, {Mazzali}, {Chornock}, {Leonard}, \& {Van Dyk}}]{Foley03}
{Foley}, R.~J., {Papenkova}, M.~S., {Swift}, B.~J., {et~al.} 2003, \pasp, 115, 1220

\bibitem[{{Ganeshalingam} {et~al.}(2012){Ganeshalingam}, {Li}, {Filippenko}, {Silverman}, {Chornock}, {Foley}, {Matheson}, {Kirshner}, {Milne}, {Calkins}, \& {Shen}}]{ejecta-trise}
{Ganeshalingam}, M., {Li}, W., {Filippenko}, A.~V., {et~al.} 2012, \apj, 751, 142

\bibitem[{{Gehrels} {et~al.}(2004){Gehrels}, {Chincarini}, {Giommi}, {Mason}, {Nousek}, {Wells}, {White}, {Barthelmy}, {Burrows}, {Cominsky}, {Hurley}, {Marshall}, {M{\'e}sz{\'a}ros}, {Roming}, {Angelini}, {Barbier}, {Belloni}, {Campana}, {Caraveo}, {Chester}, {Citterio}, {Cline}, {Cropper}, {Cummings}, {Dean}, {Feigelson}, {Fenimore}, {Frail}, {Fruchter}, {Garmire}, {Gendreau}, {Ghisellini}, {Greiner}, {Hill}, {Hunsberger}, {Krimm}, {Kulkarni}, {Kumar}, {Lebrun}, {Lloyd-Ronning}, {Markwardt}, {Mattson}, {Mushotzky}, {Norris}, {Osborne}, {Paczynski}, {Palmer}, {Park}, {Parsons}, {Paul}, {Rees}, {Reynolds}, {Rhoads}, {Sasseen}, {Schaefer}, {Short}, {Smale}, {Smith}, {Stella}, {Tagliaferri}, {Takahashi}, {Tashiro}, {Townsley}, {Tueller}, {Turner}, {Vietri}, {Voges}, {Ward}, {Willingale}, {Zerbi}, \& {Zhang}}]{Swift-3}
{Gehrels}, N., {Chincarini}, G., {Giommi}, P., {et~al.} 2004, \apj, 611, 1005

\bibitem[{{Guttman} {et~al.}(2024){Guttman}, {Shenhar}, {Sarkar}, \& {Waxman}}]{kappa-gamma}
{Guttman}, O., {Shenhar}, B., {Sarkar}, A., \& {Waxman}, E. 2024, \mnras, 533, 994

\bibitem[{{Horne}(1986)}]{Horne86}
{Horne}, K. 1986, \pasp, 98, 609

\bibitem[{{Itagaki}(2022)}]{22xlp-discovery}
{Itagaki}, K. 2022, Transient Name Server Discovery Report, 2022-2971, 1

\bibitem[{{Jha}(2017{\natexlab{a}})}]{Iax-SNe}
{Jha}, S.~W. 2017{\natexlab{a}}, in Handbook of Supernovae, ed. A.~W. {Alsabti} \& P.~{Murdin}, 375

\bibitem[{{Jha}(2017{\natexlab{b}})}]{HNB-SNe}
{Jha}, S.~W. 2017{\natexlab{b}}, in Handbook of Supernovae, ed. A.~W. {Alsabti} \& P.~{Murdin}, 375

\bibitem[{{Johnson}(1964)}]{NaID-exct-temp}
{Johnson}, H.~R. 1964, Annales d'Astrophysique, 27, 695

\bibitem[{{Kerzendorf} \& {Sim}(2014)}]{TARDIS-main}
{Kerzendorf}, W.~E. \& {Sim}, S.~A. 2014, \mnras, 440, 387

\bibitem[{{Kromer} {et~al.}(2015){Kromer}, {Ohlmann}, {Pakmor}, {Ruiter}, {Hillebrandt}, {Marquardt}, {R{\"o}pke}, {Seitenzahl}, {Sim}, \& {Taubenberger}}]{Iax-def-Kromer}
{Kromer}, M., {Ohlmann}, S.~T., {Pakmor}, R., {et~al.} 2015, \mnras, 450, 3045

\bibitem[{{Lach} {et~al.}(2022){Lach}, {Callan}, {Bubeck}, {R{\"o}pke}, {Sim}, {Schrauth}, {Ohlmann}, \& {Kromer}}]{Iax-def-Lach}
{Lach}, F., {Callan}, F.~P., {Bubeck}, D., {et~al.} 2022, \aap, 658, A179

\bibitem[{{Li} {et~al.}(2003){Li}, {Filippenko}, {Chornock}, {Berger}, {Berlind}, {Calkins}, {Challis}, {Fassnacht}, {Jha}, {Kirshner}, {Matheson}, {Sargent}, {Simcoe}, {Smith}, \& {Squires}}]{SN2002cx}
{Li}, W., {Filippenko}, A.~V., {Chornock}, R., {et~al.} 2003, \pasp, 115, 453

\bibitem[{{Long} {et~al.}(2014){Long}, {Jordan}, {van Rossum}, {Diemer}, {Graziani}, {Kessler}, {Meyer}, {Rich}, \& {Lamb}}]{Iax-def-Min}
{Long}, M., {Jordan}, George~C., I., {van Rossum}, D.~R., {et~al.} 2014, \apj, 789, 103

\bibitem[{{Magee} {et~al.}(2016){Magee}, {Kotak}, {Sim}, {Kromer}, {Rabinowitz}, {Smartt}, {Baltay}, {Campbell}, {Chen}, {Fink}, {Gal-Yam}, {Galbany}, {Hillebrandt}, {Inserra}, {Kankare}, {Le Guillou}, {Lyman}, {Maguire}, {Pakmor}, {R{\"o}pke}, {Ruiter}, {Seitenzahl}, {Sullivan}, {Valenti}, \& {Young}}]{SN2015H}
{Magee}, M.~R., {Kotak}, R., {Sim}, S.~A., {et~al.} 2016, \aap, 589, A89

\bibitem[{{McClelland} {et~al.}(2010){McClelland}, {Garnavich}, {Galbany}, {Miquel}, {Foley}, {Filippenko}, {Bassett}, {Wheeler}, {Goobar}, {Jha}, {Sako}, {Frieman}, {Sollerman}, {Vinko}, \& {Schneider}}]{Iax-relations-1}
{McClelland}, C.~M., {Garnavich}, P.~M., {Galbany}, L., {et~al.} 2010, \apj, 720, 704

\bibitem[{{Miller} \& {Stone}(1993)}]{3mShane-KAST}
{Miller}, J.~S. \& {Stone}, R. P.~S. 1993

\bibitem[{{Milne} {et~al.}(2010){Milne}, {Brown}, {Roming}, {Holland}, {Immler}, {Filippenko}, {Ganeshalingam}, {Li}, {Stritzinger}, {Phillips}, {Hicken}, {Kirshner}, {Challis}, {Mazzali}, {Schmidt}, {Bufano}, {Gehrels}, \& {Vanden Berk}}]{Iax-reddening}
{Milne}, P.~A., {Brown}, P.~J., {Roming}, P. W.~A., {et~al.} 2010, \apj, 721, 1627

\bibitem[{{Narayan} {et~al.}(2011){Narayan}, {Foley}, {Berger}, {Botticella}, {Chornock}, {Huber}, {Rest}, {Scolnic}, {Smartt}, {Valenti}, {Soderberg}, {Burgett}, {Chambers}, {Flewelling}, {Gates}, {Grav}, {Kaiser}, {Kirshner}, {Magnier}, {Morgan}, {Price}, {Riess}, {Stubbs}, {Sweeney}, {Tonry}, {Wainscoat}, {Waters}, \& {Wood-Vasey}}]{SN2009ku}
{Narayan}, G., {Foley}, R.~J., {Berger}, E., {et~al.} 2011, \apjl, 731, L11

\bibitem[{{Nicholl}(2018)}]{SuperBol}
{Nicholl}, M. 2018, Research Notes of the American Astronomical Society, 2, 230

\bibitem[{{P{\'a}l}(2012)}]{FITSH}
{P{\'a}l}, A. 2012, \mnras, 421, 1825

\bibitem[{{Phillips}(1993)}]{Phillips-rel-Phillips}
{Phillips}, M.~M. 1993, \apjl, 413, L105

\bibitem[{{Pskovskii}(1977)}]{Phillips-re-Pskovskii}
{Pskovskii}, I.~P. 1977, \sovast, 21, 675

\bibitem[{{Rodr{\'\i}guez} {et~al.}(2014){Rodr{\'\i}guez}, {Clocchiatti}, \& {Hamuy}}]{dist-NGC3938}
{Rodr{\'\i}guez}, {\'O}., {Clocchiatti}, A., \& {Hamuy}, M. 2014, \aj, 148, 107

\bibitem[{{Roming} {et~al.}(2005){Roming}, {Kennedy}, {Mason}, {Nousek}, {Ahr}, {Bingham}, {Broos}, {Carter}, {Hancock}, {Huckle}, {Hunsberger}, {Kawakami}, {Killough}, {Koch}, {McLelland}, {Smith}, {Smith}, {Soto}, {Boyd}, {Breeveld}, {Holland}, {Ivanushkina}, {Pryzby}, {Still}, \& {Stock}}]{Swift-2}
{Roming}, P. W.~A., {Kennedy}, T.~E., {Mason}, K.~O., {et~al.} 2005, \ssr, 120, 95

\bibitem[{{Siebert} {et~al.}(2020){Siebert}, {Dimitriadis}, {Polin}, \& {Foley}}]{Siebert20}
{Siebert}, M.~R., {Dimitriadis}, G., {Polin}, A., \& {Foley}, R.~J. 2020, \apjl, 900, L27

\bibitem[{{Silverman} {et~al.}(2012){Silverman}, {Foley}, {Filippenko}, {Ganeshalingam}, {Barth}, {Chornock}, {Griffith}, {Kong}, {Lee}, {Leonard}, {Matheson}, {Miller}, {Steele}, {Barris}, {Bloom}, {Cobb}, {Coil}, {Desroches}, {Gates}, {Ho}, {Jha}, {Kandrashoff}, {Li}, {Mandel}, {Modjaz}, {Moore}, {Mostardi}, {Papenkova}, {Park}, {Perley}, {Poznanski}, {Reuter}, {Scala}, {Serduke}, {Shields}, {Swift}, {Tonry}, {Van Dyk}, {Wang}, \& {Wong}}]{Silverman2012}
{Silverman}, J.~M., {Foley}, R.~J., {Filippenko}, A.~V., {et~al.} 2012, \mnras, 425, 1789

\bibitem[{{Singh} {et~al.}(2024){Singh}, {Sahu}, {Barna}, {Gangopadhyay}, {Dastidar}, {Teja}, {Misra}, {Howell}, {Wang}, {Mo}, {Yan}, {Hiramatsu}, {Pellegrino}, {Anupama}, {Joshi}, {Bostroem}, {Burke}, {McCully}, {Subramanian V}, {Li}, {Xi}, {Li}, {Li}, {Srivastav}, {Im}, \& {Dutta}}]{SN2020udy}
{Singh}, M., {Sahu}, D.~K., {Barna}, B., {et~al.} 2024, \apj, 965, 73

\bibitem[{{Srivastav} {et~al.}(2020){Srivastav}, {Smartt}, {Leloudas}, {Huber}, {Chambers}, {Malesani}, {Hjorth}, {Gillanders}, {Schultz}, {Sim}, {Auchettl}, {Fynbo}, {Gall}, {McBrien}, {Rest}, {Smith}, {Wojtak}, \& {Young}}]{SN2019gsc}
{Srivastav}, S., {Smartt}, S.~J., {Leloudas}, G., {et~al.} 2020, \apjl, 892, L24

\bibitem[{{Stritzinger} {et~al.}(2014){Stritzinger}, {Hsiao}, {Valenti}, {Taddia}, {Rivera-Thorsen}, {Leloudas}, {Maeda}, {Pastorello}, {Phillips}, {Pignata}, {Baron}, {Burns}, {Contreras}, {Folatelli}, {Hamuy}, {H{\"o}flich}, {Morrell}, {Prieto}, {Benetti}, {Campillay}, {Haislip}, {LaClutze}, {Moore}, \& {Reichart}}]{SN2008ha-SN2010ae-2}
{Stritzinger}, M.~D., {Hsiao}, E., {Valenti}, S., {et~al.} 2014, \aap, 561, A146

\bibitem[{{Stritzinger} {et~al.}(2015){Stritzinger}, {Valenti}, {Hoeflich}, {Baron}, {Phillips}, {Taddia}, {Foley}, {Hsiao}, {Jha}, {McCully}, {Pandya}, {Simon}, {Benetti}, {Brown}, {Burns}, {Campillay}, {Contreras}, {F{\"o}rster}, {Holmbo}, {Marion}, {Morrell}, \& {Pignata}}]{SN2012Z}
{Stritzinger}, M.~D., {Valenti}, S., {Hoeflich}, P., {et~al.} 2015, \aap, 573, A2

\bibitem[{{Szalai} {et~al.}(2015){Szalai}, {Vink{\'o}}, {S{\'a}rneczky}, {Tak{\'a}ts}, {Benk{\H{o}}}, {Kelemen}, {Kuli}, {Silverman}, {Marion}, \& {Wheeler}}]{SN2011ay-1}
{Szalai}, T., {Vink{\'o}}, J., {S{\'a}rneczky}, K., {et~al.} 2015, \mnras, 453, 2103

\bibitem[{{Taguchi} {et~al.}(2022){Taguchi}, {Maeda}, \& {Kawabata}}]{22xlp-classification}
{Taguchi}, K., {Maeda}, K., \& {Kawabata}, M. 2022, Transient Name Server Classification Report, 2022-2999, 1

\bibitem[{{Tomasella} {et~al.}(2016){Tomasella}, {Cappellaro}, {Benetti}, {Pastorello}, {Hsiao}, {Sand}, {Stritzinger}, {Valenti}, {McCully}, {Arcavi}, {Elias-Rosa}, {Harmanen}, {Harutyunyan}, {Hosseinzadeh}, {Howell}, {Kankare}, {Morales-Garoffolo}, {Taddia}, {Tartaglia}, {Terreran}, \& {Turatto}}]{SN14ck-Iax-relations-2}
{Tomasella}, L., {Cappellaro}, E., {Benetti}, S., {et~al.} 2016, \mnras, 459, 1018

\bibitem[{{Tonry} {et~al.}(2012){Tonry}, {Stubbs}, {Lykke}, {Doherty}, {Shivvers}, {Burgett}, {Chambers}, {Hodapp}, {Kaiser}, {Kudritzki}, {Magnier}, {Morgan}, {Price}, \& {Wainscoat}}]{SDSS-to-Bessel}
{Tonry}, J.~L., {Stubbs}, C.~W., {Lykke}, K.~R., {et~al.} 2012, \apj, 750, 99

\bibitem[{{Valenti} {et~al.}(2008){Valenti}, {Benetti}, {Cappellaro}, {Patat}, {Mazzali}, {Turatto}, {Hurley}, {Maeda}, {Gal-Yam}, {Foley}, {Filippenko}, {Pastorello}, {Challis}, {Frontera}, {Harutyunyan}, {Iye}, {Kawabata}, {Kirshner}, {Li}, {Lipkin}, {Matheson}, {Nomoto}, {Ofek}, {Ohyama}, {Pian}, {Poznanski}, {Salvo}, {Sauer}, {Schmidt}, {Soderberg}, \& {Zampieri}}]{Arnett-Valenti}
{Valenti}, S., {Benetti}, S., {Cappellaro}, E., {et~al.} 2008, \mnras, 383, 1485

\bibitem[{{Yaron} \& {Gal-Yam}(2012)}]{WISeREP}
{Yaron}, O. \& {Gal-Yam}, A. 2012, \pasp, 124, 668

\end{thebibliography}

\clearpage

\begin{appendix}
\onecolumn

\section{Multicolor photometric data reduction}
\label{appendix:BRC80-dataproc}

In this appendix, we present the detailed multicolor photometric data reduction
process used at the Baja Astronomical Observatory of the University of Szeged.
The images taken on the same night with the same color filter are averaged
to improve the signal-to-noise ratio and the limit magnitude. The instrumental
magnitudes of the SN are calculated by the image subtraction method. The
essence of this method is to subtract a pre-supernova template image of the
galaxy's environment from the supernova images. This template image is properly
scaled in size, flux, and in the full width at half maximum (FWHM) of the
star profiles' PSF with the {\tt geomap}, {\tt gregister}, {\tt psfmatch}
and {\tt linmatch} {\tt IRAF} tasks. As a result, the galaxy and the stars
in the field of view disappear, leaving only the supernova, which can then
be easily measured using aperture photometry. We used Pan-STARRS DR1 images
as templates. Finally, using the SDSS band photometric standard stars from
Pan-STARRS DR1\footnote{The Pan-STARRS DR1 archive: \url{https://outerspace.stsci.edu/display/PANSTARRS/}}
on the original unsubtracted images, the photometric standard transformation
can easily be performed. The selection of the photometry reference stars
and the calibration procedures are done in four steps. First, sources within
a 5 arcmin radius around the SN with r-band brightness between 15 and 17
mag are downloaded from the PS1 database. In the second step, we should sort
out the non-stellar sources based on the criterion $i_{PSFmag} - i_{Kronmag}
< 0.05$ for stars. To obtain reference magnitudes for  Johnson-Cousins-Bessel
B and V-band, the PS1 magnitudes are transformed into the BVRI system based
on equations and coefficients in \citep{SDSS-to-Bessel}. Finally, instrumental
magnitudes are transformed into standard magnitudes by applying a linear
color correction term (g – i) and wavelength-dependent zero points. An
atmospheric extinction correction is not necessary because the reference
stars fall close to the supernova around in a few arcminutes.

\begin{table}
\caption{Exposure time by filters used at the BRC80 telescope.}
\label{tab:BRC80-exptimes}
\begin{tabular}{ccccccc}
\hline
& B & V & g & r & i & z               \\ \hline
Exposure time [s] & 300        & 180        & 180        & 180        & 180
& 180        \\ \hline
\end{tabular}
\end{table}

\section{Multicolor photometric data}

\begin{longtable}{c c c c c c c c}
\caption{Long-term multicolor photometric follow-up data of SN 2022xlp transformed
to the standard system. Data were taken at the Baja Astronomical Observatory
of the University of Szeged with the BRC80 telescope and with the telescopes
of the LCO network.} \\
\label{tab:SN2022xlp-photometry} \\
\hline \hline
MJD & $B$ [mag] & $V$ [mag] & $g$ [mag] & $r$ [mag] & $i$ [mag] & $z$ [mag]
& Telescope \\
\hline
\endfirsthead
\caption{continued.} \\
\hline
MJD & B [mag] & V [mag] & g [mag] & r [mag] & i [mag] & z [mag] & Telescope
\\
\hline
\endhead
\hline
\endfoot
59869.11 & 16.26 $\pm$ 0.06 & 16.14 $\pm$ 0.03 & 16.13 $\pm$ 0.09 & 16.15
$\pm$ 0.02 & 16.38 $\pm$ 0.02 & 16.66 $\pm$ 0.05 & BRC80 \\
59870.04 & 16.10 $\pm$ 0.19 & 16.03 $\pm$ 0.07 & 16.00 $\pm$ 0.12 & 15.96
$\pm$ 0.16 & 16.24 $\pm$ 0.05 & 16.69 $\pm$ 0.12 & BRC80 \\
59871.11 & 16.15 $\pm$ 0.10 & 15.90 $\pm$ 0.07 & 15.87 $\pm$ 0.04 & 15.92
$\pm$ 0.03 & 16.11 $\pm$ 0.04 & 16.30 $\pm$ 0.05 & BRC80 \\
59872.12 & 16.10 $\pm$ 0.10 & 15.79 $\pm$ 0.04 & 15.79 $\pm$ 0.02 & 15.84
$\pm$ 0.02 & 16.13 $\pm$ 0.02 & 16.32 $\pm$ 0.08 & BRC80 \\
59873.13 & 15.90 $\pm$ 0.09 & 15.72 $\pm$ 0.09 & 15.77 $\pm$ 0.06 & 15.78
$\pm$ 0.07 & 16.09 $\pm$ 0.06 & 16.19 $\pm$ 0.06 & BRC80 \\
59876.5 & 16.23 $\pm$ 0.07 & 15.84 $\pm$ 0.04 & - & - & - & - & LCO \\
59879.5 & 16.69 $\pm$ 0.07 & 16.02 $\pm$ 0.04 & - & - & - & - & LCO \\
59880.26 & - & - & 16.35 $\pm$ 0.09 & 15.87 $\pm$ 0.07 & 16.03 $\pm$ 0.06
& - & LCO \\
59882.5 & 17.13 $\pm$ 0.09 & 16.20 $\pm$ 0.05 & 16.56 $\pm$ 0.13 & 15.97
$\pm$ 0.12 & 16.09 $\pm$ 0.08 & - & LCO \\
59885.51 & 17.55 $\pm$ 0.07 & - & - & - & - & - & LCO \\
59887.48 & 17.81 $\pm$ 0.07 & 16.59 $\pm$ 0.03 & 17.17 $\pm$ 0.03 & 16.28
$\pm$ 0.03 & 16.29 $\pm$ 0.03 & - & LCO \\
59890.48 & 18.05 $\pm$ 0.06 & 16.81 $\pm$ 0.03 & 17.43 $\pm$ 0.04 & 16.48
$\pm$ 0.02 & 16.47 $\pm$ 0.02 & - & LCO \\
59893.51 & 18.31 $\pm$ 0.06 & 17.00 $\pm$ 0.04 & 17.63 $\pm$ 0.04 & 16.67
$\pm$ 0.03 & 16.67 $\pm$ 0.03 & - & LCO \\
59901.22 & 18.62 $\pm$ 0.07 & 17.38 $\pm$ 0.04 & 17.92 $\pm$ 0.04 & 17.07
$\pm$ 0.03 & 17.01 $\pm$ 0.03 & - & LCO \\
59910.2 & 18.84 $\pm$ 0.07 & 17.69 $\pm$ 0.04 & 18.18 $\pm$ 0.04 & 17.43
$\pm$ 0.03 & 17.32 $\pm$ 0.03 & - & LCO \\
59915.19 & 18.92 $\pm$ 0.06 & 17.83 $\pm$ 0.05 & 18.27 $\pm$ 0.04 & 17.57
$\pm$ 0.03 & 17.46 $\pm$ 0.03 & - & LCO \\
59919.17 & - & 17.98 $\pm$ 0.20 & 18.41 $\pm$ 0.10 & 17.78 $\pm$ 0.09 & -
& - & LCO \\
59924.38 & 19.12 $\pm$ 0.09 & 18.06 $\pm$ 0.07 & 18.44 $\pm$ 0.05 & 17.82
$\pm$ 0.04 & 17.71 $\pm$ 0.05 & - & LCO \\
59926.10 & 19.28 $\pm$ 0.10 & 18.19 $\pm$ 0.05 & 18.49 $\pm$ 0.04 & 18.02
$\pm$ 0.03 & 18.04 $\pm$ 0.04 & 17.90 $\pm$ 0.10 & BRC80 \\
59931.27 & 19.21 $\pm$ 0.07 & 18.14 $\pm$ 0.05 & 18.51 $\pm$ 0.05 & 18.00
$\pm$ 0.04 & 17.86 $\pm$ 0.04 & - & LCO \\
59936.14 & 19.26 $\pm$ 0.06 & 18.26 $\pm$ 0.05 & 18.59 $\pm$ 0.05 & 18.14
$\pm$ 0.04 & 18.00 $\pm$ 0.03 & - & LCO \\
59941.33 & 19.29 $\pm$ 0.08 & 18.35 $\pm$ 0.05 & 18.66 $\pm$ 0.04 & 18.30
$\pm$ 0.05 & 18.11 $\pm$ 0.05 & - & LCO \\
59946.07 & 19.26 $\pm$ 0.08 & 18.66 $\pm$ 0.06 & 18.74 $\pm$ 0.04 & 18.54
$\pm$ 0.04 & 18.58 $\pm$ 0.05 & 18.35 $\pm$ 0.14 & BRC80 \\
59951.3 & - & 18.57 $\pm$ 0.08 & 18.95 $\pm$ 0.09 & 18.60 $\pm$ 0.07 & 18.27
$\pm$ 0.13 & - & LCO \\
59956.29 & - & - & - & - & 18.11 $\pm$ 0.21 & - & LCO \\
59961.28 & 19.65 $\pm$ 0.08 & 18.79 $\pm$ 0.06 & 18.97 $\pm$ 0.05 & 18.75
$\pm$ 0.05 & 18.51 $\pm$ 0.06 & - & LCO \\
59968.13 & 19.72 $\pm$ 0.07 & 18.92 $\pm$ 0.06 & 19.07 $\pm$ 0.04 & 18.91
$\pm$ 0.06 & 18.53 $\pm$ 0.04 & - & LCO \\
59973.42 & 19.78 $\pm$ 0.07 & 19.05 $\pm$ 0.06 & 19.23 $\pm$ 0.05 & - & -
& - & LCO \\
59975.28 & 19.57 $\pm$ 0.20 & 19.10 $\pm$ 0.10 & 19.25 $\pm$ 0.11 & - & -
& - & LCO \\
59978.23 & 20.08 $\pm$ 0.15 & 19.13 $\pm$ 0.08 & 19.28 $\pm$ 0.07 & 19.27
$\pm$ 0.07 & 18.90 $\pm$ 0.07 & - & LCO \\
59981.00 & 19.97 $\pm$ 0.20 & - & 19.55 $\pm$ 0.14 & - & - & - & BRC80 \\
59981.24 & 19.89 $\pm$ 0.19 & 19.06 $\pm$ 0.12 & 19.37 $\pm$ 0.12 & 19.41
$\pm$ 0.18 & 19.01 $\pm$ 0.13 & - & LCO \\
59981.95 & 19.80 $\pm$ 0.24 & 19.11 $\pm$ 0.16 & - & - & - & - & BRC80 \\
59982.95 & - & - & 19.68 $\pm$ 0.16 & 19.51 $\pm$ 0.13 & - & - & BRC80 \\
59985.00 & - & - & 19.53 $\pm$ 0.10 & - & - & - & BRC80 \\
59985.89 & - & - & 19.37 $\pm$ 0.09 & 19.72 $\pm$ 0.12 & 19.18 $\pm$ 0.12
& - & BRC80 \\
59986.47 & - & 19.38 $\pm$ 0.11 & 19.43 $\pm$ 0.18 & 19.44 $\pm$ 0.13 & 18.79
$\pm$ 0.25 & - & LCO \\
59991.2 & 20.36 $\pm$ 0.33 & 19.06 $\pm$ 0.17 & 19.62 $\pm$ 0.20 & 19.33
$\pm$ 0.19 & 18.93 $\pm$ 0.15 & - & LCO \\
59998.18 & 20.20 $\pm$ 0.09 & 19.40 $\pm$ 0.07 & 19.51 $\pm$ 0.06 & 19.41
$\pm$ 0.07 & 18.95 $\pm$ 0.07 & - & LCO \\
60006.35 & 20.44 $\pm$ 0.12 & 19.62 $\pm$ 0.09 & 19.89 $\pm$ 0.09 & 19.62
$\pm$ 0.09 & 19.15 $\pm$ 0.08 & - & LCO \\
60007.93 & - & - & 20.16 $\pm$ 0.21 & - & 19.21 $\pm$ 0.14 & - & BRC80 \\
60014.18 & 20.43 $\pm$ 0.11 & 19.74 $\pm$ 0.08 & 19.87 $\pm$ 0.07 & 19.68
$\pm$ 0.08 & 19.17 $\pm$ 0.07 & - & LCO \\
60014.94 & - & - & - & - & 18.93 $\pm$ 0.27 & - & BRC80 \\
60015.90 & - & - & 20.31 $\pm$ 0.15 & - & - & - & BRC80 \\
60020.92 & - & - & 20.23 $\pm$ 0.10 & - & - & - & BRC80 \\
60025.98 & 20.49 $\pm$ 0.10 & - & 20.37 $\pm$ 0.16 & - & - & - & BRC80 \\
60029.45 & 20.70 $\pm$ 0.10 & 19.96 $\pm$ 0.08 & 20.04 $\pm$ 0.06 & 19.92
$\pm$ 0.08 & 19.33 $\pm$ 0.06 & - & LCO \\
60037.08 & 20.82 $\pm$ 0.19 & 20.25 $\pm$ 0.16 & 20.32 $\pm$ 0.14 & - & -
& - & LCO \\
60045.28 & 20.91 $\pm$ 0.12 & 20.29 $\pm$ 0.12 & 20.33 $\pm$ 0.09 & 20.10
$\pm$ 0.11 & 19.42 $\pm$ 0.07 & - & LCO \\
60053.25 & 20.96 $\pm$ 0.12 & 20.23 $\pm$ 0.11 & 20.32 $\pm$ 0.09 & 20.18
$\pm$ 0.12 & 19.50 $\pm$ 0.09 & - & LCO \\
60061.2 & 21.03 $\pm$ 0.15 & 20.52 $\pm$ 0.14 & 20.48 $\pm$ 0.11 & 20.45
$\pm$ 0.14 & 19.62 $\pm$ 0.12 & - & LCO \\
\end{longtable}

\section{UV photometry}

\begin{table}
\caption{UV photometric dataset measured by the Swift Ultraviolet/Optical
Telescope (UVOT).}
\centering
\label{tab:SN2022xlp-UV-photometry}
\begin{tabular}{cccccc}
\hline
MJD & UVW2 [mag] & UVM2 [mag] & UVW1 [mag] & U [mag] & B [mag] \\ \hline
$59870.91$   & $18.507 \pm 0.212 $ & $18.951 \pm 0.248$  & $16.962 \pm 0.090$
 & $15.365 \pm 0.032$ & $15.865 \pm 0.031$ \\
$59873.24$   & $19.237 \pm 0.409$  & $23.361 \pm 16.886$ & $17.680 \pm 0.143$
 & $15.706 \pm 0.043$ & $15.824 \pm 0.032$ \\
$59876.22$   & $20.113 \pm 0.786$  & $-$                 & $18.554 \pm 0.260$
 & $16.370 \pm 0.058$ & $16.244 \pm 0.037$ \\ \hline
\end{tabular}
\end{table}

\section{The log of spectral observations}
\begin{table}
\caption{Log of the observed spectra of SN 2022xlp. The first two columns
show the date of the spectral observation in MJD and standard UTC format.
The third column displays the time since the explosion, $t_{exp}$, based
on the TARDIS fits (see Section \ref{section:spectroscopy-analysis-and-spectral-tomography}),
while the fourth and fifth column shows the phase relative to the B-band
and V-band maximum.}
\label{tab:SN2022xlp-spectra-data}
\begin{tabular}{cccccccc}
\hline
\multicolumn{1}{c}{MJD} &
\multicolumn{1}{c}{\shortstack{Date of \\ Observation}} &
\multicolumn{1}{c}{\shortstack{t$_{\mathrm{exp}}$ \\ {[}days{]}}} &
\multicolumn{1}{c}{\shortstack{Phase \\ (B max) \\ {[}days{]}}} &
\multicolumn{1}{c}{\shortstack{Phase \\ (V max) \\ {[}days{]}}} &
\multicolumn{1}{c}{\shortstack{Telescope \\ and \\ Instrument}} &
\multicolumn{1}{c}{\shortstack{Wavelength band \\ {[}$\AA${]}}} &
\multicolumn{1}{c}{\shortstack{Spectral \\ Resolution}} \\
\hline
59867.83 & 2022-10-15 19:52 & 6.0   & -4.66   & -6.23   & 3.8m-Semei/KOOLS-IFU
& 4090 - 8569  & 500     \\
59880.52 & 2022-10-28 12:28 & 19.0  & +8.03   & +6.46   & 3m-Shane/Kast 
      & 3498 - 10652 & -       \\
59887.53 & 2022-11-04 12:49 & 26.0  & +15.04  & +13.47  & 3m-Shane/Kast 
      & 3699 - 10302 & -       \\
59897.60 & 2022-11-14 14:17 & 35.8  & +25.11  & +23.54  & LCO/FLOYDS    
      & 3493 - 9939  & 400-700 \\
59897.62 & 2022-11-14 14:55 & 35.8  & +25.13  & +23.56  & LCO/FLOYDS    
      & 3491 - 9974  & 400-700 \\
59898.53 & 2022-11-15 12:36 & 37.0  & +26.04  & +24.47  & 3m-Shane/Kast 
      & 3699 - 10600 & -       \\
59904.49 & 2022-11-21 11:52 & 43.0  & +32.00  & +30.43  & 3m-Shane/Kast 
      & 3660 - 10166 & -       \\
59915.59 & 2022-12-02 13:04 & 53.8  & +43.10  & +41.53  & LCO/FLOYDS    
      & 3493 - 9968  & 400-700 \\
59923.52 & 2022-12-10 12:33 & 61.7  & +51.03  & +49.46  & LCO/FLOYDS    
      & 3492 - 9968  & 400-700 \\
59934.50 & 2022-12-21 11:53 & 73.0  & +62.01  & +60.44  & 3m-Shane/Kast 
      & 3700 - 10411 & -       \\
59937.44 & 2022-12-24 10:31 & 75.6  & +64.95  & +63.38  & LCO/FLOYDS    
      & 3771 - 9618  & 400-700 \\
59964.59 & 2023-01-20 14:16 & 102.8 & +92.10  & +90.53  & LCO/FLOYDS    
      & 3493 - 9968  & 400-700 \\
60024.43 & 2023-03-21 10:19 & 162.6 & +151.94 & +150.37 & LCO/FLOYDS    
      & 3492 - 9969  & 400-700 \\ \hline
\end{tabular}
\end{table}

\newpage
\section{Fixing the model parameters of TARDIS simulations}
\label{appendix:TARDIS-settings}
\useunder{\uline}{\ul}{}
\begin{table}
\caption{Configuration of TARDIS simulations used to fit the measured spectra
with the synthesized spectra. From top to bottom, the blocks are as follows:
\\
1) The database used for atomic and statistical physics calculations. \\
2) The TARDIS model settings: the density and abundance profile are taken
from an external data file, enabling their cell-by-cell definition. It is
important to note that while the data in this table remains constant, the
actual content of the mentioned files, i.e., the density and abundance profile,
serves as fitting parameters in the simulations. The details are discussed
in Section \ref{section:spectroscopy-analysis-and-spectral-tomography}. \\
3) Settings related to plasma calculations \\
4) Parameters for Monte Carlo iterations and the update of various physical
quantities per iteration. The quantities indexed with 'inner' refer to the
photosphere.  \\
5) The last block contains settings related to the synthesized spectrum.}
\label{tab:SN2022xlp-TARDIS-simulation-parameters-eng}
\begin{tabular}{llll}
\hline
\multicolumn{4}{c}{Constant parameters of TARDIS simulations}           
                                                                        
            \\ \hline
\multicolumn{2}{l|}{{\ul Atomic physics database:}}                     
                             & \multicolumn{2}{l}{{\ul Monte-Carlo settings:}}
               \\
{\tt atom\_data:}                              & \multicolumn{1}{l|}{kurucz\_cd23\_chianti\_H\_He.h5}
& montecarlo:                                    &               \\ \cline{1-2}
\multicolumn{2}{l|}{{\ul Model settings:}}                              
                             & \hspace{3mm} {\tt seed:}                 
     & 10000000      \\
model:                                         & \multicolumn{1}{l|}{}  
                             & \hspace{3mm} {\tt no\_of\_packets:}      
     & 50000         \\
\hspace{3mm} structure:                        & \multicolumn{1}{l|}{}  
                             & \hspace{3mm} {\tt iterations:}           
     & 20            \\
\hspace{6mm} {\tt type:}                       & \multicolumn{1}{l|}{file}
                           & \hspace{3mm} {\tt last\_no\_of\_packets:}  
   & 150000        \\
\hspace{6mm} {\tt filename:}                   & \multicolumn{1}{l|}{density.dat}
                    & \hspace{3mm} {\tt no\_of\_virtual\_packets:}   & 4
            \\
\hspace{6mm} {\tt filetype:}                   & \multicolumn{1}{l|}{simple\_ascii}
                  & \hspace{3mm} convergence\_strategy:            &    
          \\
\hspace{3mm} abundances:                       & \multicolumn{1}{l|}{}  
                             & \hspace{6mm} {\tt type:}                 
     & damped        \\
\hspace{6mm} {\tt type:}                       & \multicolumn{1}{l|}{file}
                           & \hspace{6mm} {\tt damping\_constant:}      
   & 1.0           \\
\hspace{6mm} {\tt filename:}                   & \multicolumn{1}{l|}{abundance.dat}
                  & \hspace{6mm} {\tt threshold:}                  & 0.05
         \\
\hspace{6mm} {\tt filetype:}                   & \multicolumn{1}{l|}{custom\_composition}
            & \hspace{6mm} {\tt fraction:}                   & 0.8      
    \\ \cline{1-2}
\multicolumn{2}{l|}{{\ul Plasma configuration:}}                        
                             & \hspace{6mm} {\tt hold\_iterations:}     
     & 3             \\
plasma:                                        & \multicolumn{1}{l|}{}  
                             & \hspace{6mm} {\tt t\_inner\_update\_exponent:}
& -0.5          \\
\hspace{3mm} {\tt ionization:}                 & \multicolumn{1}{l|}{nebular}
                        & \hspace{6mm} {t\_inner:}                      
&               \\
\hspace{3mm} {\tt excitation:}                 & \multicolumn{1}{l|}{dilute-lte}
                     & \hspace{8mm} {\tt damping\_constant:}          & 1.0
          \\
\hspace{3mm} {\tt radiative\_rates\_type:}        & \multicolumn{1}{l|}{dilute-blackbody}
& \hspace{8mm} {\tt threshold:}         & 0.05           \\
\hspace{3mm} {\tt line\_interaction\_type:}    & \multicolumn{1}{l|}{macroatom}
                      & \hspace{6mm} {t\_rad:}                         &
              \\
\hspace{3mm} {\tt disable\_electron\_scattering:} & \multicolumn{1}{l|}{False}
           & \hspace{8mm} {\tt damping\_constant:} & 1.0            \\
\hspace{3mm} {\tt disable\_line\_scattering:}  & \multicolumn{1}{l|}{False}
                          & \hspace{8mm} {\tt threshold:}               
  & 0.05          \\
\hspace{3mm} nlte:                             & \multicolumn{1}{l|}{}  
                             & \hspace{6mm} {w:}                        
     &               \\
\hspace{6mm} {\tt species:}                    & \multicolumn{1}{l|}{[]}
                             & \hspace{8mm} {\tt damping\_constant:}    
     & 1.0           \\
\hspace{6mm} {\tt coronal\_approximation:}     & \multicolumn{1}{l|}{False}
                          & \hspace{8mm} {\tt threshold:}               
  & 0.05          \\ \cline{3-4} 
\hspace{6mm} {\tt classical\_nebular:}         & \multicolumn{1}{l|}{False}
                          & \multicolumn{2}{l}{{\ul Spectrum settings:}}
                  \\
\hspace{3mm} continuum\_interaction:           & \multicolumn{1}{l|}{}  
                             & spectrum:                                
     &               \\
\hspace{6mm} {\tt species:}                    & \multicolumn{1}{l|}{[]}
                             & \hspace{3mm}{\tt start:}                 
     & 1700 angstrom \\
\hspace{6mm} {\tt enable\_adiabatic\_cooling:}    & \multicolumn{1}{l|}{False}
           & \hspace{3mm}{\tt stop:}               & 10000 angstrom \\
\hspace{6mm} {\tt enable\_two\_photon\_decay:} & \multicolumn{1}{l|}{False}
                          & \hspace{3mm}{\tt num:}                      
  & 600           \\
\hspace{3mm}{\tt helium\_treatment:}           & \multicolumn{1}{l|}{none}
                           & \multicolumn{2}{l}{\multirow{2}{*}{}}      
                   \\
\hspace{3mm}{\tt link\_t\_rad\_t\_electron:}   & \multicolumn{1}{l|}{0.9}
                            & \multicolumn{2}{l}{}                      
                    \\ \hline
\end{tabular}
\end{table}
\end{appendix}

\end{document}